\documentclass[prx,aps,showpacs,twocolumn,amsmath,amssymb,superscriptaddress,longbibliography]{revtex4-1}
\usepackage[english]{babel}
\usepackage{amsmath,amssymb,bbm,mathrsfs,mathtools,multirow,diagbox,xcolor,%
  graphicx,tabularx,comment,amsfonts,dsfont,lettrine,units,comment}
\usepackage{graphicx}
\usepackage[colorlinks=True,linkcolor=red,citecolor=blue,urlcolor=blue]{hyperref}
\usepackage{bookmark}
\usepackage{xcolor}
\usepackage{chngcntr}
\usepackage{braket}
\usepackage{nicefrac}
\usepackage{bm}
\usepackage{bbm}
\usepackage{braket}
\usepackage{lineno}
\usepackage{array}
\newcolumntype{C}{>{$}c<{$}}
\usepackage{newtxtext} 
\usepackage{ulem}

\begin{document}

\title{Guided accumulation of active particles by topological design of a second-order skin effect}

\author{Lucas S. Palacios}
\affiliation{Institute for Bioengineering of Catalonia (IBEC), Barcelona Institute for Science and Technology (BIST), Baldiri I Reixac 10-12, 08028 Barcelona, Spain}

\author{Serguei Tchoumakov}
\affiliation{Univ. Grenoble Alpes, CNRS, Grenoble INP, Institut N\'eel, 38000 Grenoble, France}

\author{Maria Guix}
\affiliation{Institute for Bioengineering of Catalonia (IBEC), Barcelona Institute for Science and Technology (BIST), Baldiri I Reixac 10-12, 08028 Barcelona, Spain}

\author{Ignacio Pagonabarraga}
\affiliation{Departament de F\'isica de la Mat\`eria Condensada, Universitat de Barcelona, C. Mart\'i Franqu\`es  1, 08028 Barcelona, Spain}
\affiliation{University of Barcelona Institute of Complex Systems (UBICS), Universitat de Barcelona, 08028 Barcelona, Spain}
\affiliation{CECAM, Centre Europ\'een de Calcul Atomique et Mol\'eculaire, \'Ecole  Polytechnique F\'ed\'erale de Lausanne (EPFL), Batochime, Avenue Forel 2, 1015 Lausanne, Switzerland}

\author{Samuel S\'anchez$^*$}
\email[]{ssanchez@ibecbarcelona.eu}
\affiliation{Institute for Bioengineering of Catalonia (IBEC), Barcelona Institute for Science and Technology (BIST), Baldiri I Reixac 10-12, 08028 Barcelona, Spain}
\affiliation{Instituci\'o Catalana de Recerca i Estudis Avan\c{c}ats (ICREA), Pg.  Llu\'is Companys 23, 08010 Barcelona, Spain.}

\author{Adolfo G. Grushin$^*$}
\email{adolfo.grushin@neel.cnrs.fr}
\affiliation{Univ. Grenoble Alpes, CNRS, Grenoble INP, Institut N\'eel, 38000 Grenoble, France}

\date{\today}

\maketitle


\section*{Abstract}
\textbf{
Collective guidance of out-of-equilibrium systems without using external fields is a challenge of paramount importance in active matter, ranging from bacterial colonies to swarms of self-propelled particles. 
Designing strategies to guide active matter and exploiting enhanced diffusion associated to its motion will provide insights for application from sensing, drug delivery to water remediation.
However, achieving directed motion without breaking detailed balance, for example by asymmetric topographical patterning, is challenging.
Here we engineer a two-dimensional periodic topographical design with detailed balance in its unit cell where we observe spontaneous particle edge guidance and corner accumulation of self-propelled particles.
This emergent behaviour is guaranteed by a second-order non-Hermitian skin effect, a topologically robust non-equilibrium phenomenon, that we use to dynamically break detailed balance.
Our stochastic circuit model predicts, without fitting parameters, how guidance and accumulation can be controlled and enhanced by design: a device guides particles more efficiently if the topological invariant characterizing it is non-zero.
Our work establishes a fruitful bridge between active and topological matter, and our design principles offer a blueprint to design devices that display spontaneous, robust and predictable guided motion and accumulation, guaranteed by out-of-equilibrium topology.
}

\section*{Introduction}


The last decades of research in condensed matter physics have revealed that exceptionally robust electronic motion occurs at the boundaries of a class of insulators known as topological insulators~\cite{Hasan2010,Qi2011}. 
These ideas extend beyond solid-state physics, and predict guided boundary motion in systems including photonic~\cite{Ozawa2019}, acoustic and mechanical systems~\cite{CLKane}. 

The recent discovery that topological properties emerge in the class of out-of-equilibrium systems described by non-Hermitian matrices, which includes active matter systems~\cite{Shankar20}, has opened the possibility to engineer robust behaviour out of equilibrium~\cite{Bergholtz:vn,Torres:2019jl}.
While in equilibrium topological boundary states are predicted by a non-zero bulk topological invariant, a feature known as the bulk-boundary correspondence, in non-Hermitian systems, this correspondence is broken by the skin-effect~\cite{Martinez2018b,Xiong_2018,Kunst2018,Martinez2018,Yao2018,Lee2019}. 
For example, in a one-dimensional (1D) chain of hopping particles, the first-order non-Hermitian skin effect arises from the asymmetry between left and right hopping probabilities, which results in an accumulation of a macroscopic number of modes, of the order of the system size, on one side of the system. 
This 1D effect occurs in systems without an inversion center, and has been observed in photon dynamics~\cite{Xiao:2020bda}, mechanical metamaterials~\cite{Ghatak2019,Brandenbourger:2019vg,Chen2020}, optical fibers~\cite{Weidemann311} and topoelectrical circuits~\cite{Helbig:2020bh,Hofmann20}. 
In 1D, the skin-effect occurs if a topological invariant, the integer associated to the winding of the complex spectrum of the normal modes, is non-zero~\cite{Borgnia:2020hi,Okuma20,Zhang20}.

Higher-dimensional versions of the skin effect can display a considerably richer and subtle phenomenology~\cite{Yu2020,Edvardsson2019,Liu2019,Ezawa:2019jq,Wu2021,Wu2020,Lee2019c,Li:2020fl,Ma2020,Okugawa20,Kawabata20b,Fu2020,Zhang2021} 
In this work we are interested in an elusive second-order non-Hermitian skin effect, predicted only in out-of-equilibrium systems in two dimensions (2D)~\cite{Lee2019c,Li:2020fl,Ma2020,Okugawa20,Kawabata20b,Fu2020}. 
It differs from the first-order skin effect because (i) it can occur in inversion symmetric systems, accumulating modes at opposing corners rather than edges~\cite{Hofmann20,Scheibner20}, and (ii) the number of accumulated modes is of order of the system boundary $L$, rather than its area $L^2$. 
While the first-order non-Hermitian skin effect requires inversion to be broken, e.g. due to an applied field, the emergence of the second-order non-Hermitian skin effect is guaranteed by the presence of certain symmetries~\cite{Okugawa20,Kawabata20b}.
However, predicting the second-order non-Hermitian skin effect is challenging in general, and
%
%
it remains unobserved.  
Dissipation, which drives a system out of equilibrium, is hard to control experimentally in quantum electronic devices, therefore calling for platforms to realize  the second-order non-Hermitian skin effect.
\begin{figure*}[!htb]
    \centering
    \includegraphics[width=\textwidth]{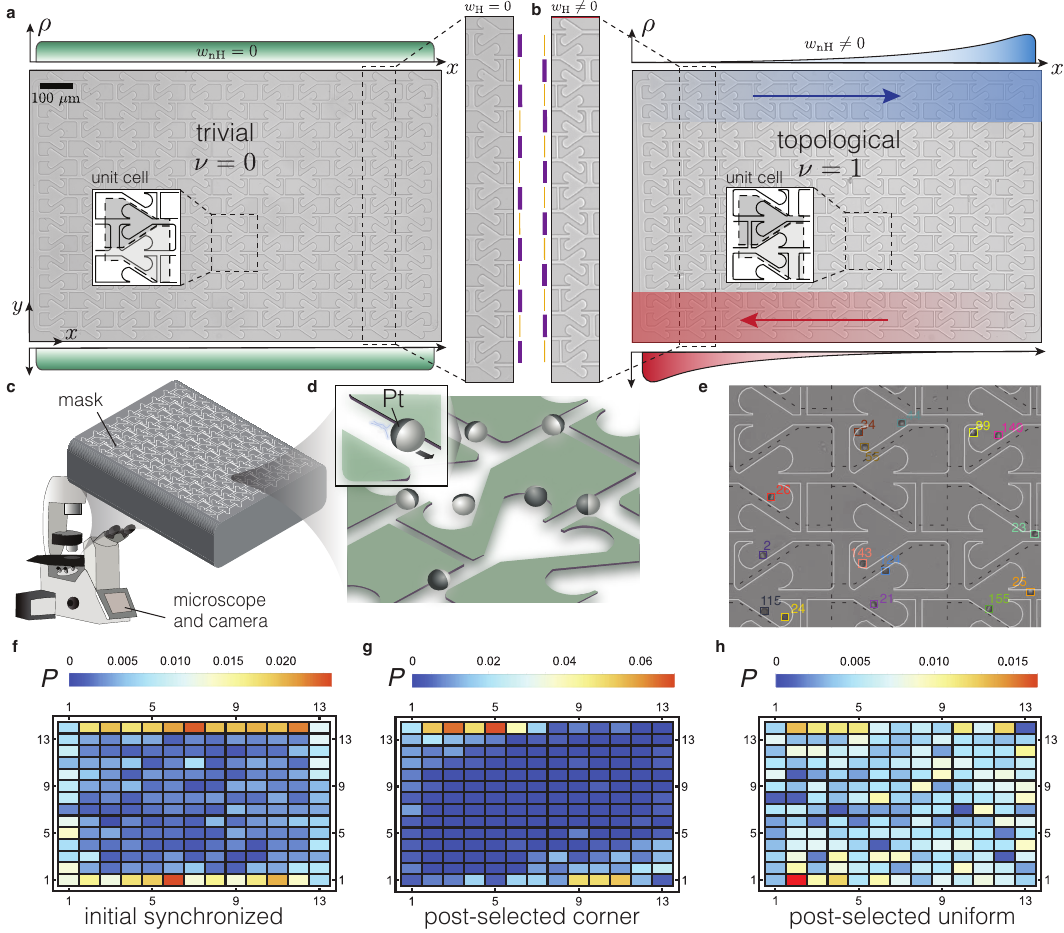}
    \caption{\label{fig:experiment}\textbf{Experimental designs.} \textbf{a} and \textbf{b} show the trivial and the topological devices respectively, with ($L_x$, $L_y$) = ($13,14$) unit cells. The insets depict the unit cells which are the same for both devices except the vertical wide and narrow channels are reversed, as shown in the zoomed central inset and emphasized with orange and purple lines. The topological device displays chiral edge modes at the top and bottom, sketched in blue and red respectively, and which are related to a non-vanishing Hermitian topological invariant, $w_{\rm H}$. These topological edge modes have a non-vanishing non-Hermitian winding number, $w_{\rm nH}$, responsible for the accumulation of active particles at the corners. This accumulation at the corners is expected to vanish for the trivial device, see the uniform density in green, because the topological invariant $\nu = w_{\rm H} w_{\rm nH}$ vanishes. \textbf{c} and \textbf{d} show a schematic of the experimental set up, where Pt-coated SiO$_2$ Janus particles self-propel when hydrogen peroxide is added, following the topographic features of each design. \textbf{e} shows a portion of the device to illustrate the tracking of particles by using a neural network, each particle is uniquely identified within our algorithm. We locate particles within the cells outlined by the dashed lines. \textbf{f}-\textbf{h} show the probability distribution data, $\mathbf{P}$, of particles on the lattice when trajectories are tracked and synchronized to start at the same initial time \textbf{f}, post-selected to start at opposing corners \textbf{g}, or post-selected to have an initial uniform distribution \textbf{h}. The particle distributions in {\bf g} and {\bf h} are shown three minutes after synchronization. These initial particle distributions are similar for both topological and trivial devices.}
\end{figure*}

Active matter systems~\cite{Bechinger:2016cf} are a natural platform to explore non-Hermitian topological physics, since these systems absorb and dissipate energy~\cite{Shankar20}. 
Often, the hydrodynamic equations that describes their flow can be mapped to a topological Hamiltonian.
This strategy predicts topologically protected motion of topological waves in active-liquid metamaterials~\cite{Souslov:2017bd,Shankar17,Souslov19}, skin-modes in active elastic media~\cite{Scheibner20}, and emergent chiral behaviour for periodic arrays of defects~\cite{Sone2019}. 
Non-Hermitian topology in active matter has been demonstrated experimentally in active nematic cells~\cite{Yamauchi2020}, and robotic~\cite{Brandenbourger:2019vg} and piezoelectric metamaterials~\cite{Chen2020}.

In this work we design microfabricated devices~\cite{simmchen2016,Teo:2016gx,Katuri:2018jv} that display a controllable second-order non-Hermitian skin effect~(see Fig.~\ref{fig:experiment}).
These devices are designed to satisfy detailed balance on their unit cell, such that the flow of particles through a unit cell vanishes.
The non-Hermitian skin effect dynamically breaks this detailed balance on the top and bottom edges, and we use it to guide and accumulate self-propelled Janus particles.
In contrast to hydrodynamic descriptions, the topological particle dynamics in our devices is quantitatively described by a stochastic circuit model~\cite{Murugan:2017ke} without fitting parameters.
It establishes that topological circuits, where a topological invariant $\nu=1$, display the second-order non-Hermitian skin effect that guides and accumulates particles more efficiently than the topologically trivial circuits, with $\nu=0$.
This phenomenon occurs without external stimuli, e.g. electrical or magnetic fields, {a useful feature for active matter applications~\cite{Bechinger:2016cf,wang2015one, gompper20202020}, and to extend our design principles to metamaterial platforms~\cite{Ozawa2019,CLKane,Lee:2018tn}.}
\begin{figure*}[!htb]
    \centering
    \includegraphics[width=\textwidth]{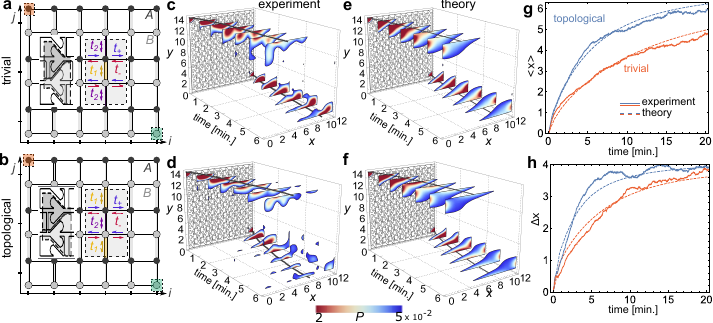}
    \caption{\textbf{Chiral edge motion.} \textbf{a, b} Illustration of the stochastic network in Eq.~\eqref{eq:master} on which we superimpose the unit cell of the experimental device (see Figs.~\ref{fig:experiment} {a},{b}). Each dot corresponds to a cell of the device and each link represents one of the transition probabilities $(t_1,t_2,t_+,t_-)$ to move in each cardinal direction (see main text). The unit cell of this lattice is the rectangle in gray, which contains black and gray nodes that are labelled by the sub-cell indexes $A,B$, respectively, introduced in Eq.~\eqref{eq:master}. \textbf{c-f} Propagation of particles starting from the top left and bottom right corners (the orange and green areas in \textbf{a, b}) for different times, for the \textbf{c, e} trivial and \textbf{d, f} topological device. We show both the \textbf{c, d} experimental and \textbf{e, f} theoretical behaviour for ($L_x$, $L_y$) = ($13,14$). The solid lines in {\bf c}-{\bf f} follow the average position of the particles starting from top left and bottom right corners. \textbf{g} Average position and \textbf{h} standard deviation in units of the cell index of a set of trajectories starting from either top left or bottom right corners of the device, which we compare with our model (see also Supplementary Discussion~1). The active particles of a topological device propagate faster than in a trivial device.}
    \label{fig:network}
\end{figure*}

\section*{Results}

\subsection*{Coupled-wire device design and stochastic model}

Our design realizes the coupled-wire construction, a theoretical tool to construct topological phases~\cite{Meng19}. 
This is possible by the precise engineering of microchannel devices (see Methods)~\cite{simmchen2016}.
Each device contains two types of horizontal microchannels, the wires, which are coupled vertically, forming a 2D mask (see Fig.~\ref{fig:experiment}{a}, {b}). 
The horizontal microchannels are consecutive left or right oriented hearth-shaped ratchets, that favour a unidirectional motion towards their tip~\cite{Katuri:2018jv}. 
Their left-right orientation alternates vertically. 
These horizontal microchannels are coupled by vertical microchannels that are straight, designed to imprint a symmetric  vertical motion to the nanoparticles. 
The vertical microchannels alternate in width, with successive narrow and wide channels. Wide channels are more likely to be followed by the active particles than narrow channels.

Using these principles, we design two types of devices that we coin trivial and topological, depicted in Figs.~\ref{fig:experiment}a and b, respectively. 
They only differ in that the narrow and wide channels exchange their roles along the vertical direction (see central inset of Figs.~\ref{fig:experiment}a and {b}). 
Both trivial and topological designs have the same number of left and right oriented ratchets, so they satisfy global balance.

Active particles are injected within the device and move along the microchannels walls, see Figs.~\ref{fig:experiment}{c} and {d} and Methods. 
We track the particles from recorded videos with an in-house developed software based on a tiny YOLOv3 neuronal network~\cite{yolov3}. 
We locate the particles within a grid of regularly spaced cells (dashed lines in Fig.~\ref{fig:experiment}{e}). 
We use the recorded trajectories as a database, which we can post-select and synchronize in order to study the average particle motion starting from various initial configurations (see Methods). 
In Figs.~\ref{fig:experiment}{f, g} and {h} we show different initial distributions of particles for latter use.
Fig.~\ref{fig:experiment}{f} shows the particle distributions when all recorded trajectories are synchronized to start at the same initial time. 
We observe more particles at the borders than in the bulk because of the particle flux from outside the device. 
Since this flux cannot be controlled, we cannot directly choose the initial particle distribution. 
We overcome this limitation by post-selecting trajectories so that they start either from the top left and bottom right cells (Fig.~\ref{fig:experiment}{g}), or from a uniform distribution over all cells (Fig.~\ref{fig:experiment}{h}). 
This post-selection is possible due to our tracking system, and it is a differentiating aspect between active particle and electronic systems.

We study devices built out of ($L_x,L_y$) = ($12,6$) and ($L_x,L_y$) = ($13,14$) unit-cells in the horizontal ($x$) and vertical ($y$) directions, with either a small or a large density of injected active particles (see Methods).
In the main text, we focus on the larger device, with ($L_x,L_y$) = ($13,14$), see Fig.~\ref{fig:experiment}. 
All other devices and the corresponding results are shown in the Supplementary Discussion~2.

Because of the irregular microchannel walls and the collisions between particles, we model the collective motion of active particles as a Brownian motion in a stochastic network with transition probabilities between the cells introduced in Fig.~\ref{fig:experiment}e. 
Cells are represented by nodes in our model, depicted in Figs.~\ref{fig:network}a,b. 
Neglecting correlations between particles due to collisions, that we refer to as {\it jamming}, the continuous-time Markov master equation governs the probability distribution of the particles in time as~\cite{Murugan:2017ke}
\begin{linenomath}\begin{align}
    \tau \frac{\partial {\bf P}}{dt} = \hat{\bm W}\cdot {\bf P},
    \label{eq:master}
\end{align}\end{linenomath}
where ${\bf P} = P_{ij \sigma}$ is the probability to observe a particle on site $(\sigma,i,j)$, $\sigma \in \{A,B\}$
is the sub-cell index and $(i,j)$ are the primitive-cell indices (see Figs.~\ref{fig:network}a and b). 
$\tau$ is the timescale for a particle to move between adjacent sites. 
$\hat{\bm W}$ is the transition rate matrix (written explicitly in the Supplementary Discussion~1).
It depends on the four transitions probabilities: $t_{\pm}$ for the motion along or against the ratchet-like microchannels and, $t_1$ and $t_2$ for the motion along the wide and narrow microchannels, respectively. 
The time scale $\tau$ and the conditional probabilities are extracted from our experimental data in Supplementary Discussion~1. $\tau$ is roughly ten seconds and since it depends on various factors, from the concentration of chemicals to illumination, in practice we compute its probability distribution for each experiment. Also, we obtain $(t_1, t_2, t_+, t_-) \approx (0.13,0.20,0.52,0.15)$.
The large ratio $t_+/t_-$ shows the motion is unidirectional on the horizontal axis and $t_2/t_1 \approx 1.5 \neq 1$ confirms that we can explore the difference between trivial and topological devices.

\subsection*{Topological chiral edge motion}

\begin{figure}
    \centering
    \includegraphics[width=\columnwidth]{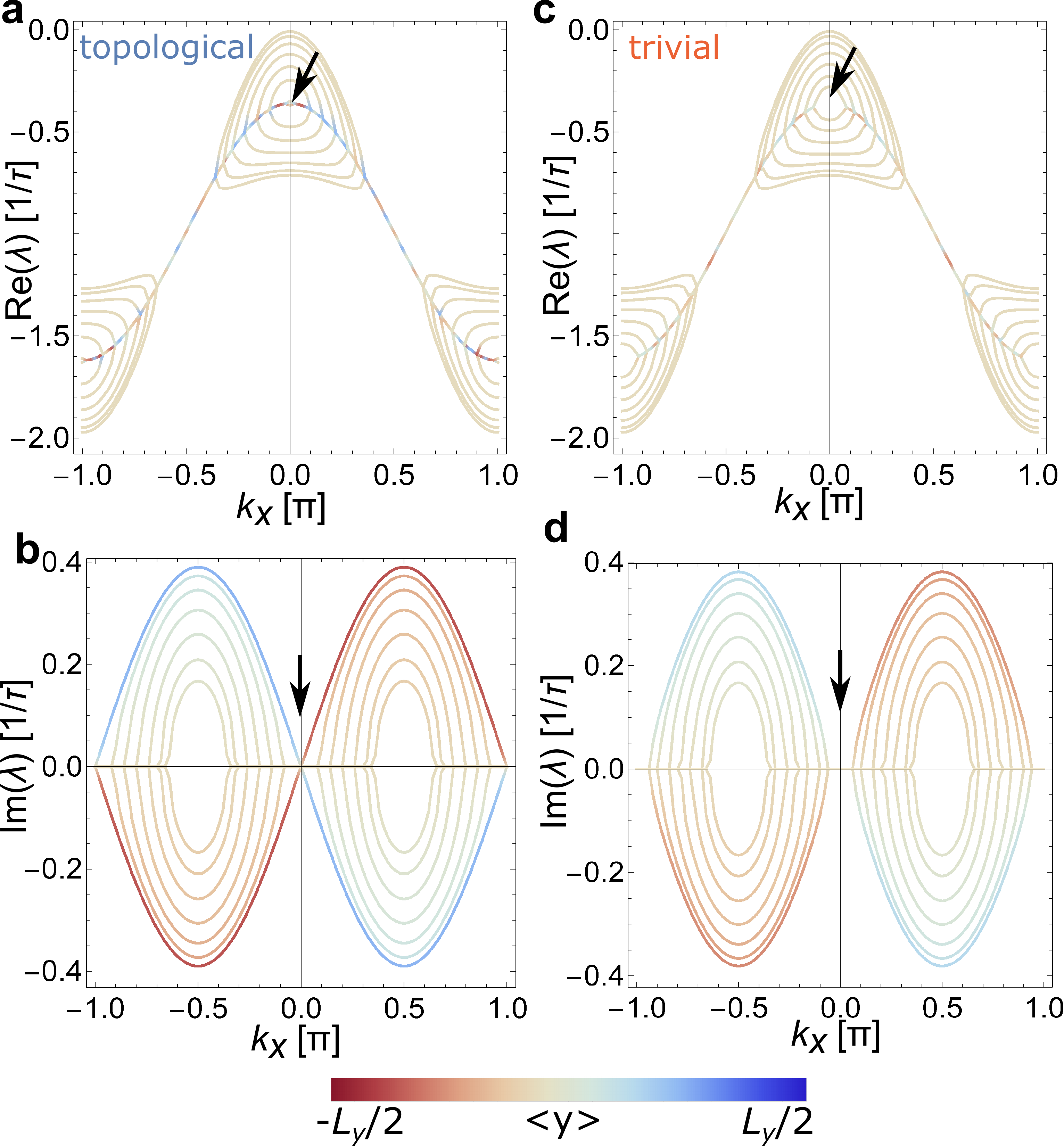}
    \caption{\textbf{Stochastic chiral edge states.} Real \textbf{a},\textbf{c} and \textbf{b},\textbf{d} imaginary part of the normal modes of the rate matrix $\hat{\bm W}$ for open boundary conditions in the $y$ direction, for the \textbf{a},\textbf{b} topological and \textbf{c},\textbf{d} trivial devices.  The color denotes the average $\langle y \rangle$ position of a normal mode. Modes in green are delocalized and correspond to bulk states. Modes in blue and red correspond are strongly localized on the top or bottom edges, respectively. The arrows in {\bf c},{\bf d} highlight the presence or absence of topological edge modes {close to $\mathbf{k}=0$, where their lifetime $\mathrm{Re}(\lambda)$ is largest.}}
    \label{fig:hermitian}
\end{figure}

\begin{figure*}[!htb]
    \centering
    \includegraphics[width=\textwidth]{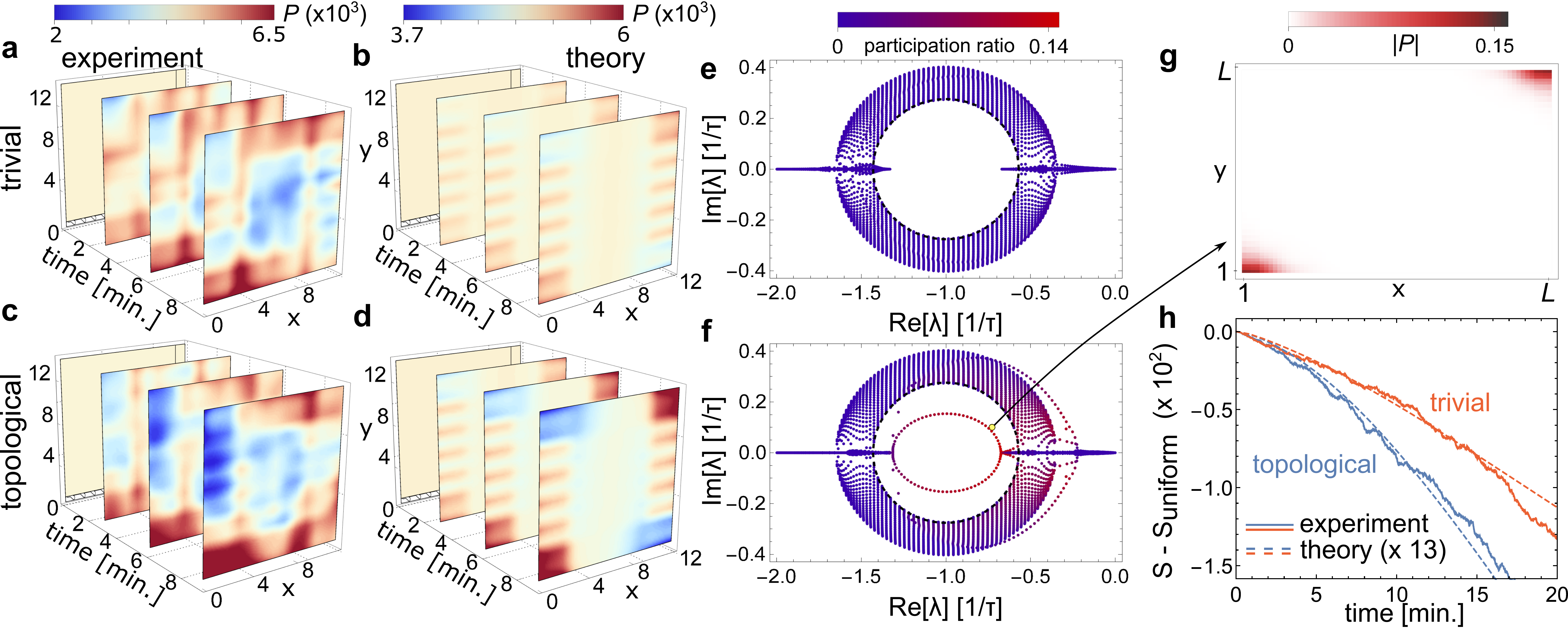}
    \caption{\textbf{Second-order non-Hermitian skin effect of the active particles.} \textbf{a-d} We compare the distribution of particles in the trivial \textbf{a,c} and topological \textbf{b,d} devices with ($L_x$, $L_y$) = ($13,14$), observed experimentally \textbf{a,b} and predicted theoretically \textbf{c,d}. We observe that more particles locate on the top right and bottom left corners. \textbf{e,f} Parametric representation of the real and imaginary parts of the spectrum of normal modes for the model of the trivial \textbf{e} and topological \textbf{f} devices with $(L_x,L_y) = (L,L) = (70,70)$, colored with the participation ratio of each normal mode $\sum_{\sigma  ij} |P_{\sigma ij}|^4$, which is small for a delocalized mode. The gap in the periodic band structure is within the dashed circle (see Supplementary Discussion~1). The number of localized modes within the point gap is proportional to the size of the device, $L$, and are localized at the corners as shown in \textbf{g}. \textbf{h} Shannon entropy of the particle distribution over time. The smaller entropy in the topological device is associated to an accumulation of particles at the corners, signaling the second-order non-Hermitian skin effect. The theoretical figures match the experimental curves when multiplied by a factor $\times 13$, suggesting that particle jamming, that occurs frequently for higher densities of active particles, contributes to enhance the accumulation.}
    \label{fig:skin}
\end{figure*}

We first use the post-selection technique to explore the edge dynamics of an ensemble of trajectories synchronized to start either at the top left or the bottom right corners (see Fig.~\ref{fig:experiment}g). 
We select 281 trajectories, of about 17 minutes each, for the trivial device, and 327 trajectories, of about 16 minutes each, for the topological case.

In Figs.~\ref{fig:network}c and d we show the density of active particles as a function of time for trivial (c) and topological (d) devices. 
Qualitatively, the particle distribution propagates unidirectionally along the edge, faster in the topological device than in the trivial one. 
Quantitatively, the time-dependent average displacement $\langle x \rangle$ and spread $\Delta x \equiv \sqrt{\langle (x - \langle x \rangle)^2\rangle}$ for particles starting from the top left corner, confirms this behaviour, see Figs.~\ref{fig:network}{g} and {h}. 
We observe that active particles in the topological device are ahead of the trivial device, by about one unit cell after three minutes. 

The motion of active particles is understood decomposing Eq.~\eqref{eq:master} into the normal modes, ${\bf P}_{\bf k}$, defined by
\begin{linenomath}\begin{align}
    \hat{\bm W}\cdot {\bf P}_{\bf k} = \lambda_{\bf k} {\bf P}_{\bf k},
    \label{eq:normal}
\end{align}\end{linenomath}
where $\lambda_{\bf k}$ is a complex scalar which depends on wavevector ${\bf k} = (k_x,k_y)$ for periodic lattices. 
The real part of $\lambda_{\bf k}$ sets the lifetime of the normal mode, and its curvature $\partial_k^2 {\rm Re}(\lambda_{\bf k})$ at ${\bf k} = 0$ sets its diffusion coefficient. 
The slope of the imaginary part $\partial_k {\rm Im}(\lambda_{\bf k})$ at ${\bf k} = 0$ sets the velocity of the normal mode (see Supplementary Discussion~1). 
For open boundaries along $y$ and periodic along $x$, the spectrum $\lambda_{\bf k}$ can be represented as a function of $k_x$. 
The boundary condition is such that the probability distribution vanishes outside the lattice.
The resulting spectrum is shown in Figs.~\ref{fig:hermitian} a,b for the topological device and c,d for the trivial one. 
The spectrum is colored according to the localization of the normal modes, where red and blue colors denote states at the top and bottom edges, respectively. The normal modes localized at the edge govern the propagation of a particle distribution localized at an edge.
Some edge modes have a chiral group velocity, shown by the slope of the imaginary part of the spectrum at $k_x = 0$ in Fig.~\ref{fig:hermitian}b, and are absent for the trivial device (see Figs.~\ref{fig:hermitian}c,d). 
If in addition to enforcing a vanishing probability distribution outside the lattice we impose that detailed balance is preserved at the boundary, an additional edge potential partially hybridizes the edge modes with bulk modes, but does not remove them (see Supplementary Discussion~1). 
%

The existence of edge modes is guaranteed by a bulk Hermitian topological invariant. 
In the vertical direction the couplings, $t_1$ and $t_2$, alternate between weak and strong (see Fig. \ref{fig:experiment}b). 
This is the stochastic equivalent of the Su-Schrieffer-Heeger model of the polyacetylene chain~\cite{Su79}, which is characterized by a winding number $w_{\mathrm{H}}$ in the $y$ direction (see Supplementary Discussion~1).
$w_{\mathrm{H}}=1$ and $w_{\mathrm{H}}=0$ for the topological and trivial device models, respectively.
This implies that they respectively have, or not, topological edge modes for a boundary in the $y$ direction, indicated by black arrows in Figs.~\ref{fig:hermitian}b, d.

The topological edge modes have a noticeable effect on the edge particle dynamics of the active particles.
We probe them by initializing Eq.~\eqref{eq:master} with a probability distribution localized at the top left corner. 
We compare the experimental results with theoretical predictions in Figs.~\ref{fig:network}e-h, using the parameters $t_1,t_2,t_+,t_-$ and $\tau$ set by our statistical analysis of the experimental data.
The theoretical curves qualitatively reproduce the experimental trends without any fitting parameter, for both topological and trivial devices. 
This demonstrates that the displacement in the topological device is larger as a consequence of the topological edge modes.

The above observations are reproduced for smaller devices and larger densities of active particles, see Supplementary Discussion~2. 
We find that for larger densities, the particles are slower than theory predicts, an effect we attribute to particle jamming, neglected in our model. 
A lower density of particles prevents jamming, in which case the motion compares better with our model.

\subsection*{Corner accumulation from second-order non-Hermitian skin effect}

The detailed balance of the unit cell implies that $\hat{\bm W}$ has no strong topological invariant~\cite{Kawabata19} (see Supplementary Discussion~1).
Moreover, $\hat{\bm W}$ has inversion symmetry, implying that the first-order skin-effect vanishes.

To derive the second-order skin effect of $\hat{\bm W}$ we consider the topological edge modes, ${\bf P}_{s,\chi}$, at the top ($\chi = +$) and bottom ($\chi = -$) described by a 1D equation $\tau \partial_t {\bf P}_{s,\pm} = H_{s,\pm} {\bf P}_{s,\pm}$ (see Supplementary Discussion~1).
%
These modes locate on either $A$ (for $\chi = +$) or $B$ (for $\chi = -$) sub-lattice, have a ballistic propagation $\langle x \rangle_{\pm} = \pm(t_+ - t_-) t/\tau$, and diffuse by an amount $\Delta x = \sqrt{(t_+ + t_-)t/\tau}$ after a time $t$ (see Supplementary Discussion~1). 
The real part of $H_{s,\pm}$ 
is finite and indicates that the contribution of the topological edge modes to the total probability decays over a timescale $\tau_d = \tau/(t_1+t_2)$. 
$\tau_d$ sets how far the particles propagate due to the topological edge modes.

The edge modes $H_{s,\pm}$ have a finite 1D winding number $w_{\mathrm{nH}}=\pm1$ that implies a 1D non-Hermitian skin effect~\cite{Borgnia:2020hi,Okuma20,Zhang20}.
Since the edge modes are spatially separated, active particles can accumulate at the top and bottom corners. 
We detect the accumulation of active particles experimentally by post-selecting trajectories that start from a uniform configuration (see Fig.~\ref{fig:experiment} h. and Methods). 
We observe an accumulation of active particles at the corners which is larger in the topological device (Figs.~\ref{fig:skin} a,c) and that qualitatively compares with our model (Fig.~\ref{fig:skin} b,d). 

This observation can be made quantitative using the Shannon entropy of the particle distribution
$S = -\sum_{ij\sigma} P_{ij \sigma} \ln\left( P_{ij \sigma} \right)$.
The entropy is maximal for a uniform distribution of particles, $S < S_{\rm uniform} = \ln(L_x L_y)$, and it decreases if particles localize. 
We average out other sources of particle localization unrelated to the non-Hermitian skin-effect by averaging the probability distribution over neighboring cells (see Supplementary Discussion~2). 
The experimental and theoretical entropies are depicted in Fig.~\ref{fig:skin} h.
Both figures show a smaller entropy in the topological device than in the trivial one. 
They depart from each other at the same rate, yet the absolute values of the experimental entropies are a factor 13 smaller than theory. 
Our model thus captures the difference between trivial and topological devices, but underestimates the accumulation that occurs in the experiment. 
A potential reason is that, in the experiment, a local increase in the number of active particles leads to particle jamming, neglected in our model. 
When we decrease the density of particles to reduce jamming, the entropy still drops but is similar for topological and trivial devices. 
This suggests that there is a critical density of particles to observe corner accumulation.

The corner accumulation we observe is a consequence of the chiral motion of topological edge modes, that accumulate at the corners for long times. 
In the topological device the chiral edge motion occurs because $w_{\mathrm{H}}\neq 0$, and the corner accumulation occurs because $w_{\mathrm{nH}}\neq0$ for $H_{s,\pm}$. 
The two devices thus have a different topological invariant
\begin{linenomath}\begin{equation}
    \label{eq:invariant}
    \nu = w_{\mathrm{H}}w_{\mathrm{nH}},
\end{equation}\end{linenomath}
which equals one or zero for the topological and trivial devices, respectively.
The topological invariant \eqref{eq:invariant} highlights the coupled-wire nature of our system, and is reminiscent of weak topological phases. This observation motivates different viewpoints on the higher-order topological behaviour we observe, as we discuss in the Supplementary Discussion 1.G.
In the Supplementary Discussion~1.F we show that this invariant is equivalent to that in Ref.~\cite{Okugawa20}, and that it signals a second-order skin-effect as follows. 
First, $\hat{\bm W}$ has inversion symmetry, and a point gap spectrum in which corner modes appear only for open-boundary conditions in both directions (Figs.~\ref{fig:skin} f, g). 
Second, the number of corner modes scales as the edge length $L$, rather than the system size $L^2$, a defining characteristic of the second-order the skin-effect~\cite{Kawabata20b}.

Our work establishes a strategy to design circuits that spontaneously break detailed balance to guide and accumulate active matter, enforced by robust out-of-equilibrium topological phenomena. 
This design strategy and the $3$D printing capabilities, permit to envision 3D extensions where the accumulation of particles is guaranteed by non-planar surfaces (e. g. using ramps, or different levels) \cite{Yang:2018ji} or by implementing dynamic topographical pathways topology \cite{Aschenbrenner:2019ui}.
\\

\section*{Methods}
\label{methods}

\subsection{Particle preparation}
\label{met:particles}

Crystal slides of $25 \times 25$ mm are sequentially cleaned with acetone and isopropanol in a sonication bath for two minutes. The glass slides are then dried with compressed air and treated with oxygen plasma for 10 minutes. Commercial silica microparticles (Sigma-Aldrich) of $5~\mu$m diameter size are deposited on the glass slides by drop-casting and left to dry at room temperature. We then sputter a $10$ nm layer of Pt (Leica EM ACE600) to integrate the catalytic layer on the silica microparticles. The samples are kept in a dry and closed environment. The Janus particles are released for each experiment after being briefly exposed to an Ar-plasma in order to increase their mobility. For each experiment, we dilute particles from a third of a glass slide in $1$~mL of water via sonication for a few seconds, after what they are ready to be used.

\subsection{Microfabricated model substrate for topological guidance: design and fabrication}
\label{met:microchannels}

The microscale features that define the masks are designed and microfabricated on a silicon wafer, and later transferred by replication to a thin structure of polydimethylsiloxane (PDMS). The microchannel design is created with a computer-aided design (CAD) software (AutoCAD, Autodesk) and is made of ratchet-like structures that favor directional trajectories of the active particles~\cite{Katuri:2018jv}. Direct writing laser lithography (DWL 66FS, Heidelberg Instruments) is used to produce an AZ$^{\tiny \circledR}$ resist master (AZ 1512HS, Microchemicals GmbH) of $1.5~\mu$m thickness on the silicon wafer. A PDMS replica from the rigid mold/master is produced to obtain the open channel microfluidic device with the desired features. For reusability purposes, the master is silanized with trichloro (1H,1H,2H,2H-perfluorooctyl)sylane (Sigma-Aldrich) by vapor phase for one hour at room temperature, this operation reduces the adhesion of PDMS to the substrate. To obtain a PDMS thin layer with the inverse pattern from the rigid mold (so called PDMS replica), the PDMS (Sylgard 184, Dow Corning) monomer and cross-linker are mixed at a ratio of 10:1 and degassed for one hour. Afterwards, the solution is spin-coated at $1000$ rpm for $10$ seconds on the top of the master and cured for $4$ hours at $65^{\circ}$C. The replica are carefully released, obtaining the desired open microchannel designs with the sub-micrometre step-like topographic features that allow the topological guidance of active particles.

\subsection{Setup fabrication}

A circular well of $2$ mm height and $1.5$ cm diameter is designed with a CAD software and post-treated with Slic3r and Repetier softwares in order to generate the G-code required for 3D printing it with a Cellink's Inkredible+ 3D printer. Polydimethylsiloxane (PDMS) SE1700 cross-linker and monomer are mixed at a ratio of $1$:$20$, and the viscous solution is added to the cartridge used in the 3D printer. The pneumatic extrusion of the 3D printer allows to print the PDMS-based structures on clean crystal slides ($60 \times 24$ mm) at a pressure of $200$ kPa. Once the well structure is printed, the slides are left overnight at $65^{\circ}$C for curing, to obtain a stable structure. To ensure good sealing of the 3D-printed well, a mixture of 1:10 PDMS Sylgard 184 is added around the outer side of the well, followed by a curing period of four hours at $65^{\circ}$~C.

The glass slides with the 3D-printed wells are sequentially cleaned with acetone and isopropanol, and dried with compressed air. The glass substrate is activated by exposure to oxygen plasma for $30$ seconds. The microratchet-like patterns in Sec.~\ref{met:microchannels} are then immediately transferred to the wells to ensure good attachment of the patterned PDMS to the glass. The whole setup is ready and it is kept in a dry and closed environment until use. Before use, the setup is cleaned with a sequential wash of acetone and isopropanol, and exposed to oxygen plasma for $30$ seconds. The setup design is such that it minimizes the accumulation of gas bubbles at the center of the sample which permits long-time experiments, from $30$ minutes to one hour.

\subsection{Active particles concentration}
\label{met:setup}

We perform our experiment for two different concentrations of active particles. This allows to compare how the density of active particles qualitatively affects our experiment. Indeed, a larger density of particles enhances the Brownian motion, due to the many collisions, but also decreases the velocity of particles and leads to more clusters. The two densities are obtained by injecting particles in either one of the two solutions:
\begin{enumerate}
    \item {\it high density of active particles.} We dilute $17.5~\mu$L of the solution of Janus particles prepared as detailed in the Methods describing particle preparation with $35~\mu$L of H$_2$O$_2$ at  2\% per volume and $47.5~\mu$L of water.
    \item {\it low density of active particles.} We first dilute $300~\mu$L of the solution of Janus particles prepared as detailed in the Methods describing particle preparation with $700~\mu$L of water. Then, we dilute $17.5~\mu$L of this solution with $35~\mu$L of H$_2$O$_2$ at  2\% per volume and $47.5~\mu$L of water.
\end{enumerate}
This way we expect a ratio of concentrations of a third to a half between high and low densities of active particles.

\subsection{Tracking}

In order to track a large amount of particles, we have developed a Python script based on a neuronal network detection algorithm~\cite{yolov3}. This neuronal network is taught to recognize a particle from a video frame and to discard anything else. Once trained, we use this neuronal network to detect particles on every frame of the video. The experiments are recorded using a contrast microscope (Leica) with a 10x objective at 2 frames per second. A particle fits in a box of $8$-$9$~px$^2$ on the resulting video. Our software relates the detected particles between frames to find the trajectory of a given particle. The following paragraphs detail the procedure to train the neuronal network and to construct trajectories from the particle we detect on each frame.

To train the neuronal network, we construct a dataset of images that show one or many particles. We perform this step manually by cropping a square image from a video frame. This cropped image is centered close to a particle's center, keeping the selected particle within, and is three times the size of a particle. We also draw a square region of the size of the particle on this cropped image, like the squared, coloured, boxes in Fig.~\ref{fig:experiment}e, that indicate the location of the particle on the cropped image. We construct our dataset by applying this procedure on many particles. We extend this dataset by also adding cropped images that do not contain any particle. This way, our dataset is composed of 8700 images with particles and 8700 images that do not show any particle. We input this dataset to a neuronal network based on tiny YOLOv3~\cite{yolov3}. We simplify the network by transforming the images to 8-bit grayscale and considering there are only two sizes for the boxes surrounding the particles. We train the network for a day and obtain a  mAP-50 larger than $95\%$, which is sufficient for our purpose. This generates a configuration file that we use to detect particles of an image. We illustrate the output of this detection algorithm in Fig.~\ref{fig:experiment}e, where an identifier is assigned to each detected particle.

To build the trajectory of a particle located at position ${\bf r}_t$ at time $t$, we estimate that its location ${\bf r}_{t+1}$ at a time $t+1$ is
\begin{equation}
\label{eq:tracking}
    {\bf r}_{t+1} = {\bf r}_{t} + \frac{{\bf r}_{t}-{\bf r}_{t-1}}{2},
\end{equation}
where the second term is an estimate of the displacement of the particle, which is zero for the first frame since we miss the position ${\bf r}_{-1}$. We use this position estimate to crop a square image of the frame $t+1$ centered on ${\bf r}_{t+1}$ and twice the size of a particle. We apply our neuronal network detection algorithm on this cropped image to detect all particles it may contain. We calculate the distance between each detected particle and each of the particles we are tracking at time $t$, keeping always the lowest distance calculated and the identification number (ID) of the associated particle. We create a list with these IDs and distances. First, we filter by selecting only those distances that have the same id as the particle we want to detect. If any, we will chose as ${\bf r}_{t+1}$ the closer object detected. If we cannot find the same ID, we select $r_{t+1}$ as the closest object detected to this cropped image. If finally, we have not detected any particle, we set ${\bf r}_{t+1} = {\bf r}_t$. We apply this algorithm for a set of particles we manually select at the initial frame, as well as particles that may enter into the frame while recording. This tracking procedure allows to track particles even when they collide; only clogs of $5$ particles or more may result in particle loss. Particles may leave the sample; if this happens their position is fixed to that frame and is removed after a period of inactivity. Also particles can adhere electrostatically to the PDMS, or to each other, or be inactive. Therefore we end the trajectory if the particle is inactive for a long period of time. We use GPU acceleration for the neuronal network detection, based on the latest NVIDIA™ drivers.

The filtered trajectories are given in $x$,$y$ pixel coordinates, which we map to the cell coordinate of our model. This transformation is obtained after aligning the axes of the recorded videos and decomposing the image over the lattice structure in Fig.~\ref{fig:experiment} \textbf{e}, shown with dashed lines. We then assign each of the now aligned ($x$,$y$) coordinate to a cell index, using the Python library \textit{shapely}.

\subsection{Post-selection of trajectories}
\label{met:postselect}

We organize the recorded trajectories in a database for each experimental configuration. We only keep track of the position in the unit-cell index to eliminate any intrinsic motion of a particle within a cell and thus to reduce noise. In the manuscript we consider two post-selected configurations, (1) where particles start at the top left and bottom right corners (see Fig.~\ref{fig:experiment}g) and, (2) where particles start from a uniform distribution (see Fig.~\ref{fig:experiment}h) over the lattice. The two post-selection procedures are as follows:
\begin{enumerate}
     \item {\bf Post-selected corner.} We select particles starting from the corners in two steps. We first select trajectories that go through one of the two corners. Then, among the selected trajectories, we remove all frames before the one where the particles first enters one of the two corners. This ensemble of trajectories is what we use for the analysis in Figs.~\ref{fig:network} e-h.
    \item {\bf Post-selected uniform.} We select particles uniformly scattered over the sample based on simulated annealing. We focus on the probability distribution $P_{\sigma ij}(t = 0)$ of all trajectories on their first frame. We scan trajectories from the longest to the shortest and remove the first frames of a trajectory until the newly generated distribution of particles $P'_{\sigma ij}(t = 0)$ is more uniform than the one,  $P_{\sigma ij}(t = 0)$, at the previous step. We consider the distribution is more uniform when 
    \begin{linenomath}\begin{align}
        C(P') = \sqrt{ \sum_{\sigma ij} \left(P'_{\sigma ij}(t = 0) - \frac{1}{L_x L_y}\right)^2 } < C(P).
    \end{align}\end{linenomath}
    During this procedure, we remove trajectories that have less than $6$ minutes of video to remove noise in the first frames. This ensemble of trajectories is what we use for the analysis in Fig.~\ref{fig:skin}.
\end{enumerate}

\section*{Data availability} 

Supplementary Information is available for this paper. The data that supports our finding is available on Zenodo~\cite{dataset}. Additional data are available upon request.\\

\section*{Code availability}
The code used to generate the figures is available upon request.


%

\textbf{Acknowledgements.} 
A. G. G and S. T. are grateful to M. Brzezinska, M. Denner, T. Neupert, Q. Marsal, S. Sayyad, and T. S\'{e}pulcre for discussions.
A. G. G and S. T acknowledge financial support from the European Union Horizon 2020 research and innovation program under grant agreement No 829044 (SCHINES). A. G. G is also supported by the ANR under the grant ANR-18-CE30-0001-01 (TOPODRIVE). L. P. is grateful to J. Katuri for discussions about ratchet design and to J. Fuentes for the PDMS wells fabrication. L.P acknowledges financial support from MINECO for the FPI BES-2016-077705 fellowship. M.G. thanks MINECO for the Juan de la Cierva fellowship (IJCI2016-30451), the Beatriu de Pin\'{o}s Programme (2018-BP-00305) and the Ministry of Business and Knowledge of the Government of Catalonia. I.P. acknowledges support from Ministerio de Ciencia, Innovaci\'{o}n
y Universidades (Grant No. PGC2018-098373-B-100), DURSI
(Grant No. 2017 SGR 884), and SNF (Project No. 200021-
175719). S.S. acknowledges the CERCA program by the Generalitat de Catalunya, the Secretaria d'Universitats i Recerca del Departament d'Empresa i Coneixement de la Generalitat de Catalunya through the project 2017 SGR 1148 and Ministerio de Ciencia, Innovaci\'{o}n y Universidades (MCIU) / Agencia Estatal de Investigaci\'{o}n (AEI) / Fondo Europeo de Desarrollo Regional (FEDER, UE) through the project RTI2018-098164-B-I00. S. S. acknowledge financial support from  the European Research Council (ERC) under the European Union's Horizon 2020 research and innovation programme (grant agreement No 866348). All the authors acknowledge MicroFabSpace and Microscopy Characterization Facility, Unit 7 of ICTS “NANBIOSIS” from CIBER-BBN at IBEC for their support in the masks design and fabrication. \\

\textbf{Author Contribution.} 
M. G. designed the microchannel designs and PDMS wells. L. P. fabricated the microchannel devices, performed the experimental work and extracted the corresponding data, developing also a the neural-network based tracking system to evaluate the trajectories of the active Janus particles. L. P., S. T. and A. G. G. analyzed the experimental data, with input from I. P. and S. S. S. T. derived the theoretical model and computed its observables. 
S. T. and A. G. G. wrote the manuscript, with input from all authors. A. G. G. devised the initial concepts for the experimental setup and for the theoretical modeling, and supervised the project. \\

\textbf{Competing Interests.}
The authors declare that they have no competing interests.\\

\clearpage
\newpage

\onecolumngrid
\renewcommand{\figurename}{Supplementary Figure}
\setcounter{figure}{0} 

\begin{center}
    \large{\bf Supplementary Information: Guided accumulation of active particles by topological design of a second-order skin effect}
\end{center}

\section*{Supplementary Figures}

\begin{figure*}[h]
    \centering
    \includegraphics[width=\textwidth]{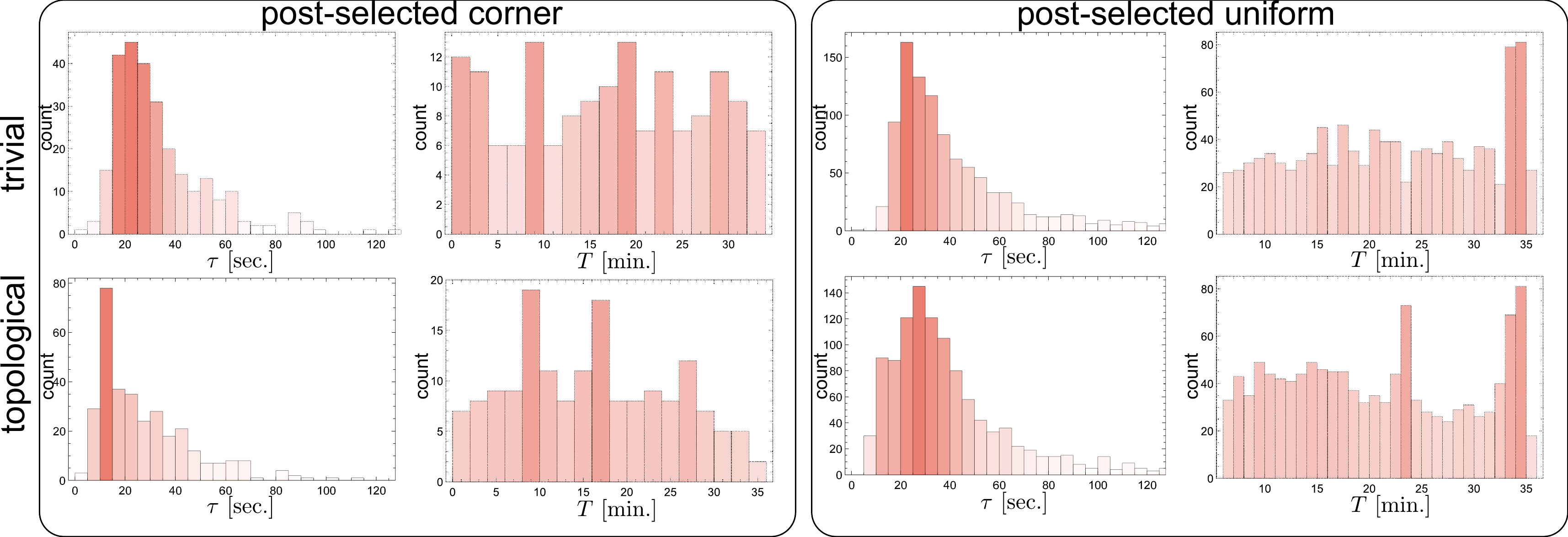}
    \caption{Distribution of characteristic times for each selection of particles for the devices with $(L_x,L_y) = (13,14)$ and a small density of active particles, for both topological and trivial devices. The timescale $\tau$ corresponds to the average time for a particle to go from one cell to the next. The timescale $T$ corresponds to the total length of a trajectory.}
    \label{fig:times}
\end{figure*}

\begin{figure*}[h]
    \centering
    \includegraphics[width=\columnwidth]{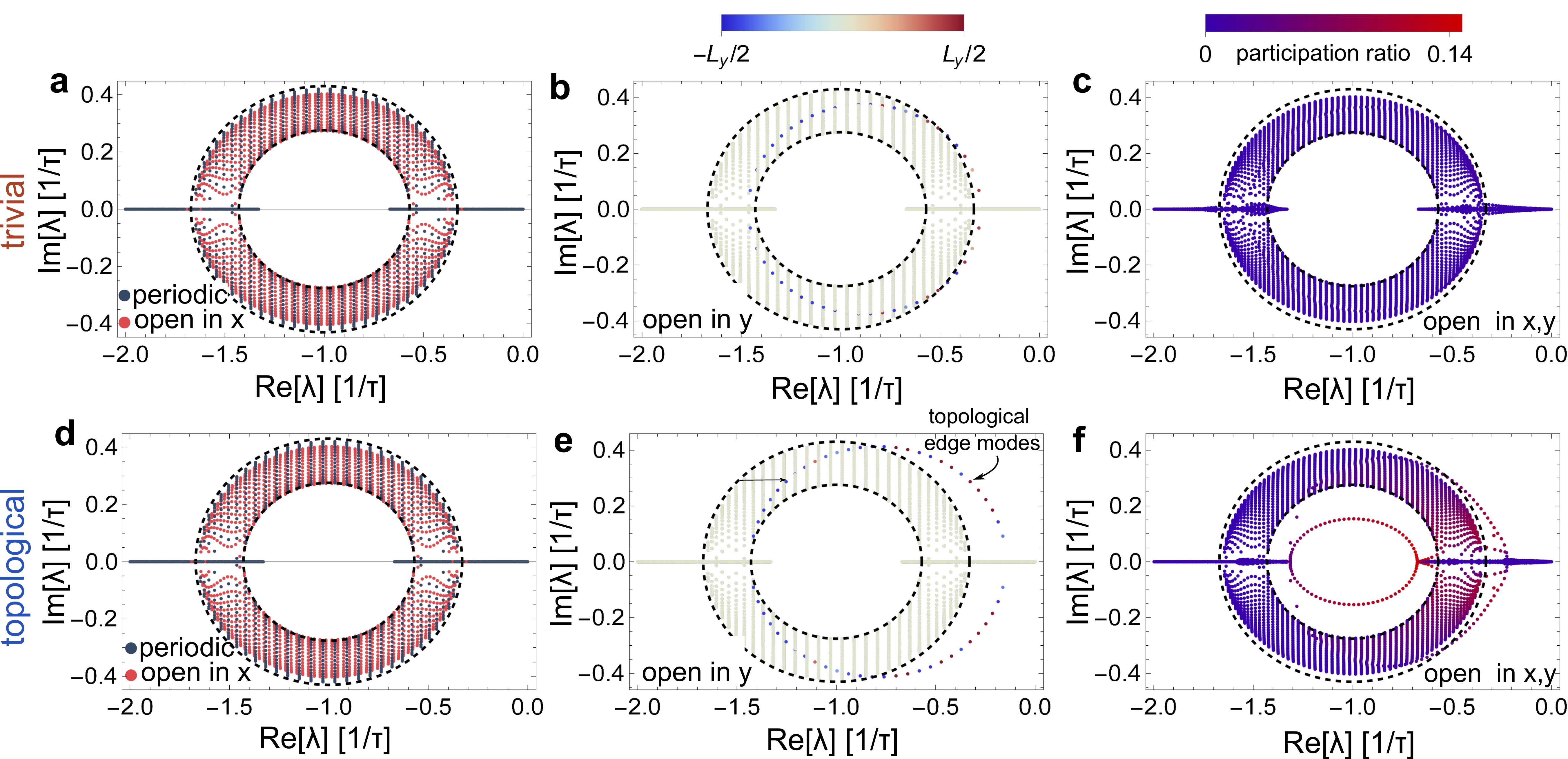}
    \caption{Parametric representation of the real and imaginary part of the spectrum of the normal modes for the trivial \textbf{a}-\textbf{c} and topological \textbf{d}-\textbf{f} devices, for ($L_x$,$L_y$)=($L$,$L$) = ($70$,$70$) lattice sites. {\bf a,d} Spectrum for periodic boundary conditions in both directions (in blue) superimposed with that for open boundary conditions in $x$ (in red). The two spectra coincide and indicate no edge mode. {\bf b,e} Spectrum for open boundary conditions in $y$. The points are colored with respect to the average position $\langle y \rangle$ of the normal mode. The topological edge modes~\eqref{eq:SS2} partly hybridize with bulk modes, this is seen as a gap opening for larger values of ${\rm Re}[\lambda]$. {\bf c,f} Spectrum for open boundary conditions in $x$ and $y$. The points are colored with respect to the participation ratio $\sum_{\sigma ij}|P_{\sigma ij}|^4$ of the normal mode, a quantity which is small for delocalized modes. As shown in the main text, the modes in the inner dashed circle are localized at the top right and bottom left corners and are related to the second order non-Hermitian skin-effect.}
    \label{fig:boundaries}
\end{figure*}

\begin{figure*}[h]
    \centering
    \includegraphics[width=0.40\columnwidth]{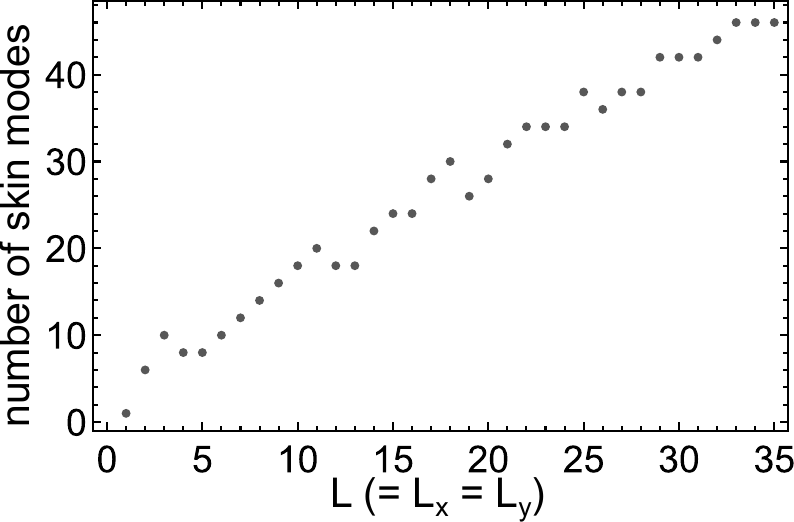}
    \caption{Number of skin modes for open boundary conditions in both $x$ and $y$ directions for the topological device, as a function of system size $L$ where $(L_x,L_y) = (L,L)$. This number corresponds to the number of states for open boundary conditions within the point gap found with periodic boundary conditions (i.e. within the inner circle in Supplementary Fig.~\ref{fig:boundaries}f). The number of skin modes increases linearly with the perimeter, which is characteristic of a second-order non-Hermitian skin effect~\cite{Kawabata20b}.}
    \label{fig:linear}
\end{figure*}

\begin{figure*}[h]
    \centering
    \includegraphics[width=\textwidth]{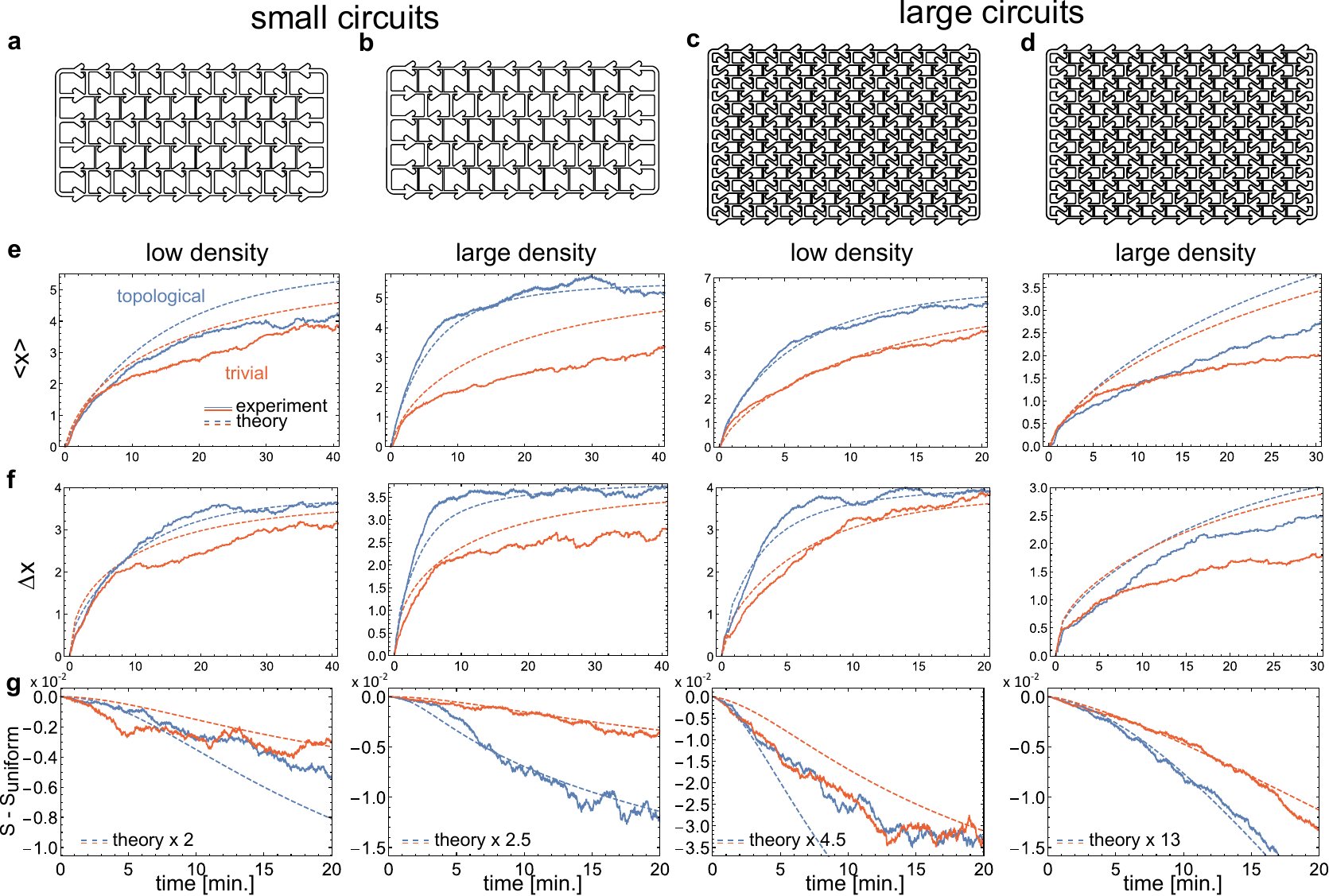}
    \caption{{\bf a},{\bf b},{\bf c},{\bf d} Depiction of the microfluidic devices used in our experiments. The trivial {\bf a} and topological {\bf b} small, $(L_x,L_y) = (12,6)$, designs, and the trivial {\bf c} and topological {\bf d} large, $(L_x,L_y) = (13,14)$, designs. In the main text we focus on the results obtained for the designs {\bf c} and {\bf d}. We consider four experimental situations with small ($L_x,L_y$) = ($12,6$) and large ($L_x,L_y$) = ($13,14$) microfluidic devices, and with low and large density of active particles. A larger density of particles enhances the Brownian motion, but also decreasest the velocity of particles and leads to more clusters. {\bf e},{\bf f} Complementing Figs.~2{g}, {h}, we show the the time evolution of the average position and spread for all devices, when tracking particles initially at the top left corner (see Methods, post-selection of trajectories). We compare these quantities for the trivial and topological devices between experiments and theory. The model parameters are assigned separately for each case, using the experimental data (see Section~\ref{sec:expvalues}), and there are no fitting parameters. {\bf g} Complementing Figs.~4{ e}, we show the time evolution of the entropy when tracking particles initially spread uniformly over the device (see Methods, post-selection of trajectories).}
    \label{fig:review}
\end{figure*}

\newpage
\clearpage
\section*{Supplementary Discussion 1 : Stochastic model of the device\label{app:theory}}

The design of the device where the Janus particles move in is inspired by the coupled-wire construction \cite{Kane02,Meng19,PhysRevB.93.195136}. In this construction one-dimensional wires are coupled vertically to construct topological phases. Within each device we describe the motion of active particles as a random walk described by the continuous-time Markov master equations
\begin{align}
    \left\{
    \begin{array}{r}
        \tau \frac{dP_{A,ij}}{dt} = t_1(P_{B,i,j+1} - P_{A,ij}) + t_2(P_{B,ij} - P_{A,ij}) + t_{+} P_{A,i-1,j} - t_{-} P_{A,ij} + t_{-} P_{A,i+1,j} - t_+ P_{A,ij},\\
        \tau \frac{dP_{B,ij}}{dt} = t_1(P_{A,i,j-1} - P_{B,ij}) + t_2(P_{A,ij} - P_{B,ij}) + t_{+} P_{B,i+1,j} - t_{-} P_{B,ij} + t_{-} P_{B,i-1,j} - t_+ P_{B,ij},
    \end{array}
    \right.
    \label{eq:kolm}
\end{align}
where the probability distribution, $P_{\sigma,ij}$, is decomposed over the discrete coordinates of the network, with $\sigma \in (A,B)$ the sub-cell indexes (see Figs.~2 a,b) and $(i,j) \in \mathbb{N}^2$ are the unit cell coordinates. This equation is balanced to ensure probability conservation, $\sum_{\sigma ij} P_{\sigma,ij} = 1$, and also, since the equation is irreducible, it has a unique stationary probability distribution with $dP_{\rm st.}/dt = 0$~\cite{DasbiswasE9031}. It is convenient to introduce the transition matrix $\hat{\bm W}$ such that
\begin{align}
    \tau \frac{d{\bf P}}{dt} = \hat{\bm W} {\bf P},
\end{align}
where $({\bf P})_{\sigma ij} = P_{\sigma,ij}$. Eq.~\eqref{eq:kolm} has the form of the Schr\"{o}dinger equation with a non-Hermitian Hamiltonian. 

\subsection{Experimental values}
\label{sec:expvalues}

We fix the values of the parameters in Eq.~\eqref{eq:kolm} by counting the number of times we see a particle moving along the four types of bulk links in the experiment. We have performed the experiment for two device sizes, $(L_x,L_y) = (12,6)$ and $(L_x,L_y) = (13,14)$ (see Figs.~\ref{fig:review} a-d) and, for a low and a high density of active particles (see Methods, Experimental Setup).

For each situation we obtain the following transition probabilities
\begin{enumerate}
    \item For our largest device, with $(L_x,L_y) = (13,14)$, and a low density of active particles:
    \begin{itemize}
        \item $( t_1, t_2, t_+, t_- ) =  ( 0.154, 0.212, 0.512, 0.122 )$ for the trivial device,
        \item $( t_1, t_2, t_+, t_- ) = (  0.214, 0.128, 0.545, 0.112 )$ for the topological device.
    \end{itemize}
    These are the values we use to compare our experiment with the model in Figs.~3 c,d.
    \item For our largest device, with $(L_x,L_y) = (13,14)$, and a high density of active particles:
    \begin{itemize}
        \item $( t_1, t_2, t_+, t_- ) =  ( 0.126, 0.206, 0.461, 0.207 )$ for the trivial device,
        \item $( t_1, t_2, t_+, t_- ) = (  0.203, 0.119, 0.466, 0.212 )$ for the topological device.
    \end{itemize}
    These are the values we use to compare our experiment with the model in Figs.~4 e.
    \item For our smallest device, with $(L_x,L_y) = (12,6)$, and a low density of active particles:
    \begin{itemize}
        \item $( t_1, t_2, t_+, t_- ) =  ( 0.150, 0.199, 0.519, 0.132 )$ for the trivial device,
        \item $( t_1, t_2, t_+, t_- ) = ( 0.179, 0.123, 0.579, 0.119 )$ for the topological device.
    \end{itemize}
    \item For our smallest device, with $(L_x,L_y) = (12,6)$, and a high density of active particles:
    \begin{itemize}
        \item $( t_1, t_2, t_+, t_- ) =  ( 0.143, 0.191, 0.514, 0.152 )$ for the trivial device,
        \item $( t_1, t_2, t_+, t_- ) = (  0.215, 0.125, 0.541, 0.120 )$ for the topological device.
    \end{itemize}
\end{enumerate}
We observe an asymmetry between vertical and horizontal motions since $t_1 + t_2 = 0.33 < 0.67 = t_+ + t_-$; the motion along the horizontal axis is easier because the ratchets are aligned while vertical micro-channels are not. This asymmetry in the design is a consequence of the spatial constraints during device fabrication and it also helps to distinguish better topological and trivial devices in the experiment. Indeed, the contribution of topological edge modes to the displacement of particles is largest for an asymmetric network with $t_1 + t_2 < t_+ + t_-$, because it increases the decay time $\tau_d = \tau/(t_1+t_2)$ of the edge modes. This condition is satisfied in the present experiment because the ratchets favour the horizontal motion of the Janus particles.
Also, one can expect to observe a ballistic regime after a time 
\begin{align}
    \tau_b = (t_+ + t_-)\tau/(t_+ - t_-)^2,
\end{align}
after which $\langle x\rangle > \Delta x$. Since in our experiment $\tau_b/\tau_d = 1.3 > 1$, it allows us to observe the ballistic regime.

Another parameter that enters our model is the typical time, $\tau$, for a particle to move from one cell to another. We evaluate this time separately for each active particle by dividing the total time of its trajectory by the number of times it goes from one cell to another. This time is different for each active particles for a variety of reasons, for example because of the differences in particle sizes or local chemical environment. We evaluate the probability distribution of $\tau$, $\mathcal{P}(\tau)$, for each initial configuration we pick up (see Fig~\ref{fig:times}). When we compare our model with the experiment we average all the quantities over the probability distribution of $\tau$. For example, for the average position we compute
\begin{align}
    \langle \langle x \rangle \rangle = \sum_{\tau} \langle x \rangle \mathcal{P}(\tau),
\end{align}
where the average position $\langle x \rangle = \sum_{\sigma ij} x_{\sigma ij} P_{\sigma ij}$ is evaluated with the probability distribution of our model, in Eq.~\eqref{eq:kolm}, for a given value of $\tau$. In the text we omit the double bracket notation but this averaging procedure is always performed.

\subsection{Bulk solution}
\label{app:bulk}
In the situation of a infinite or periodic lattice, we can decompose the solution over the basis of Bloch solutions such that $P_{\sigma ij} = \sum_{k_x , k_y} P_{\sigma}({\bf k}) e^{i(k_x i + k_y j)}$ where $\{k_x,k_y\} = \{n\pi a/L,m\pi a/L\}$ with $a$ the lattice spacing, $L$ the lattice size and $\{n,m\} \in \mathbb{N}^2$. In this basis Eq.~\eqref{eq:kolm} can be written for each ${\bf k}$ independently, $\tau d{\bf P_{\bf k}}/dt = \hat{\bm W}_{\bf k} {\bf P_{\bf k}}$, with
\begin{align}
    \tau \frac{d}{dt} \left( 
        \begin{array}{c}
            P_A({\bf k},t)\\
            P_{B}({\bf k},t)
        \end{array}
    \right) &= \left( 
        \begin{array}{cc}
            -1 + t_+ e^{-ik_x} + t_- e^{ik_x} & t_1 e^{ik_y} + t_2\\
            t_1 e^{-ik_y} + t_2 & -1 + t_+ e^{ik_x} + t_- e^{-ik_x}
        \end{array}
    \right)
    \left( 
        \begin{array}{c}
            P_A({\bf k},t)\\
            P_{B}({\bf k},t)
        \end{array}
    \right),
    \label{eq:kolmbloch}
\end{align}
which we can diagonalize to write
\begin{align}
    \tau \partial_{t} P_{\eta}({\bf k},t) = \lambda_{\eta}({\bf k}) P_{\eta}({\bf k},t),
    \label{eq:eigenP}
\end{align}
where $\eta = \pm$ denotes the two eigensolutions with normal mode $P_{\eta}$ and eigenvalue $\lambda_{\eta}$. The eigenvalues are
\begin{align}
    \label{eq:bulkspectrum}
    \lambda_{\eta} = -1 + (t_+ + t_-)\cos(k_x) \pm \sqrt{(t_2+t_1 \cos(k_y))^2 + t_1^2 \sin^2(k_y) - (t_--t_+)^2\sin^2(k_x)}.
\end{align}
The time evolution of the normal modes is then $P_{\eta}({\bf k},t) = P_{\eta}({\bf k}) e^{\lambda_{\eta \bf k}t}$, so if we define the distribution at $t = 0$ by 
\begin{align}
    P({\bf x},t = 0) = \sum_{\bf k} \sum_{\eta} c_{\eta}({\bf k}) P_{\eta}({\bf k}) e^{i(k_x x + k_y y)},
\end{align}
then the probability distribution at a time $t$ is
\begin{align}
    P({\bf x},t) = \sum_{\bf k} \sum_{\eta} c_{\eta}({\bf k}) P_{\eta}({\bf k}) e^{\lambda_{\eta \bf k}t + i(k_x x + k_y y)}.
\end{align}
Since ${\rm Re}(\lambda_{\eta \bf k}) \leq 0$, with a maximum at ${\bf k} = 0$ with ${\rm Re}(\lambda_{\eta \bf k}) = 0$, the infinite or periodic lattice tends towards the uniform distribution in space at long times.

\subsection{Real and imaginary eigenvalues}

The imaginary and real parts of the eigenvalues $\lambda_n$ determine the average position and variance of the active particles. Indeed, for an initial probability distribution that is uniform and that we decompose over the eigensolution, $P({\bf x},t = 0) = \sum_{\bf k} \sum_{\eta} c_{\eta}({\bf k}) P_{\eta}({\bf k}) e^{i(k_x x + k_y y)}$, we have
\begin{align}
    &\langle {r}_{i} \rangle_t = \sum_{\bf x} r_{i} P({\bf r},t) = -\sum_{\eta = \pm} c_{\eta}({\bf k} = 0) \partial_{k_{i}} {\rm Im}\left[ \lambda_{\eta}({\bf k} = 0) \right] t,\\
    &\Delta_t r_{i}^2 = \langle {r}_{i}^2 - \langle r_{i} \rangle_t^2 \rangle_t = -\sum_{\eta = \pm} c_{\eta}({\bf k} = 0) \partial_{k_{i}}^2 {\rm Re}\left[ \lambda_{\eta}({\bf k} = 0) \right] t,
    \label{eq:avgstd}
\end{align}
where $\mathbf{r} = (x,y)$ is the position over cells (see Fig.~1 e). These two relations show that for an initially uniform distribution, only the eigenvalues at ${\bf k} = 0$ matter. As explained in the main text, the slope of the imaginary part is related to the ballistic motion and the curvature of the real part is related to diffusion.

\subsection{Edge modes}
\label{app:edgemodes}

In this section we derive the spectrum of edge modes for open boundary conditions in the $y$ direction, and periodic in $x$. Because the lattice is periodic in the $x$ direction, we can still apply the procedure in Sec.~\ref{app:bulk} for the horizontal ($x$) direction. This leads to the equation
\begin{align}
    \tau \frac{d}{dt} \left( 
        \begin{array}{c}
            P_{A,j}(k_x,t)\\
            P_{B,j}(k_x,t)
        \end{array}
    \right) &= \left( 
        \begin{array}{cc}
            -1 + t_+ e^{-ik_x} + t_- e^{ik_x} & t_2\\
            t_2 & -1 + t_+ e^{ik_x} + t_- e^{-ik_x}
        \end{array}
    \right)
    \left( 
        \begin{array}{c}
            P_{A,j}(k_x,t)\\
            P_{B,j}(k_x,t)
        \end{array}
    \right)
    +
    t_1 \left( 
        \begin{array}{c}
            P_{B,j+1}(k_x,t)\\
            P_{A,j-1}(k_x,t)
        \end{array}
    \right)
    ,
    \label{eq:kolmblochX}
\end{align}
where $j$ labels the lattice site in the vertical ($y$) direction. Since the bulk spectrum~\eqref{eq:bulkspectrum} is invariant by the transformation $k_y \rightarrow -k_y$, the solutions for open boundaries can be expanded as
\begin{align}
    \label{eq:expansionSS}
    P_{A,j} &= a_+ e^{i k_y j} + a_- e^{-i k_y j} = A_1 \cos({k_y j}) + A_2 \sin(k_y j),\\
    P_{B,j} &= b_+ e^{i k_y j} + b_- e^{-i k_y j} = B_1 \cos({k_y j}) + B_2 \sin({k_y j}).
\end{align}
The value of $k_y$ is not necessarily real and is set by the boundary condition. Also, if we write $k_y = \chi + i\mu$, we have that
\begin{align}
    \cos(k_y j) &= \cos(\xi j) \cosh(\mu j) - i \sin(\xi j) \sinh(\mu j),\\
    \sin(k_y j) &= \sin(\xi j) \cosh(\mu j) + i \cos(\xi j) \sinh(\mu j),
\end{align}
and when replacing these expressions in Eq.~\eqref{eq:bulkspectrum} we have to check that the real part of $\lambda_n$ is negative, so the solutions stay normalized. We consider two types of open boundary conditions in  $y$: (1) one that preserves chiral symmetry and, (2) one that preserves detailed balance. 

As we show next, the boundary condition (1) is a useful approximation; its chiral symmetry allows us to show in a simple way why the edge states have a non-hermitian skin-effect in the next section. The price to pay is that we do not recover the exact numerical edge spectrum, but only an approximation to it. The boundary condition (2) is more involved analytically, but recovers the exact numerical edge spectrum with the same topological information, which is discussed in the next section. \\

(1) {\bf Boundary condition with chiral symmetry.} A way to describe open boundary conditions is to solve Eq.~\eqref{eq:kolm} with the constraints $P_{B, L_y + 1} = 0$ and $P_{A, 0} = 0$. These constraints prevent a particle from moving away from the region $y\in[1,L_y]$. The constraint $P_{B, L_y + 1} = 0$ implies that
\begin{align}
    P_{B,j} = B_1 \frac{\sin(k_y(L_y + 1 - j))}{\sin(k_y(L_y+1))}.
\end{align}
This is input in the equation for $P_{A,0}$ in Eq.~\eqref{eq:kolmblochX} to give the consistency relation
\begin{align}
    \label{eq:consistency1}
    \frac{t_1}{t_2} = - \frac{\sin(k_y(L_y +1))}{\sin(k_y L_y)} \xrightarrow[Ly\to\infty]{k_y = \pi + i \mu} e^{\mu}.
\end{align}
This transcendental equation has $L_y$ \emph{real} solutions for $k_y \in [0,\pi)$ if only $t_2/t_1 < 1$, this corresponds to the trivial device which has no edge mode. If $t_1/t_2 > 1$, as in the topological device, this equation has only $L_y-1$ real solutions for $k_y \in [0,\pi)$. The missing real solution is that at $k_y = \pi$. Note that it would be missing at $k_y = 0$ if $t_1/t_2 < -1$, but this does not occur since conditional probabilities are positive. This missing mode is actually substituted by an evanescent mode, with $k_y = \pi + i \mu$, solution to Eq.~\eqref{eq:consistency1}. In the limit where $L_y$ is large, the first term in the square-root in Eq.~\eqref{eq:bulkspectrum} vanishes and the eigenvalues are
\begin{align}
    \label{eq:lambdaSChi}
    \lambda_{\chi = \pm} = -1 + (t_+ + t_-)\cos(k_x) + \chi i(t_--t_+)\sin(k_x).
\end{align}
They correspond to the top ($\chi = +$) and bottom ($\chi = -$) interface, respectively. Compared to Eq.\eqref{eq:lambdaSChi} the numerical edge modes seen in Supplementary Fig.~\ref{fig:boundaries}e are shifted by a constant. To obtain the exact spectrum we discuss the second type of boundary conditions, that preserve detailed balance.\\~~

(2) {\bf Boundary condition with detailed balance.} The previous boundary condition does not respect detailed balance because it removes the hopping term to outside the lattice without tuning-off the associated on-site sink. This on-site sink is the $-1 = -(t_1 + t_2 + t_+ + t_-)$ term in \eqref{eq:kolmbloch}. Imposing detailed balance at the boundary leads to the following condition at the $j = L_y$ boundary,
\begin{align}
    \label{eq:boundA}
    \lambda P_{A,L_y} &= (-1 + t_1 + t_{+} e^{-ik_x} + t_- e^{ik_x}) P_{A,L_y} + t_2 P_{B,L_y}.
\end{align}
This equation looks similar to the previous one, where we impose $P_{B,L_y+1} = 0$, but with an additional $t_1 P_{A,L_y}$ contribution. This boundary condition can be worked out with~\eqref{eq:kolm} and~\eqref{eq:bulkspectrum} to give
\begin{align}
    \label{eq:boundary2}
    \left[ 2 \cos(k_y) - \underbrace{\frac{t_2}{\lambda - (-1 + t_1 + t_{+} e^{-ik_x} + t_- e^{ik_x})}}_{\equiv z} \right] P_{B,L_y} = P_{B,L_y-1}.
\end{align}
The second term in the bracket would be zero in the boundary condition (1). We combine~\eqref{eq:boundary2} with Eq~\eqref{eq:expansionSS} and get
\begin{align}
    \label{eq:Pb2}
    P_{B,j} = B_1 \left( \cos(k j) - \frac{\cos(k (L_y + 1)) - z \cos( k L_y )}{\sin(k (L_y + 1)) - z \sin( k L_y )} \sin(k j) \right).
\end{align}
We can then insert this in the equation for the $j = 1$ boundary, leading to
\begin{align}
    \lambda P_{B,1} = ( -1 + t_1 + t_+ e^{ik_x} + t_- e^{-ik_x}) P_{B,1} + t_2 P_{A,1},
\end{align}
and that can be transformed to
\begin{align}
    \label{eq:consistency2}
    \frac{1}{z} = 2 \cos(k_y) - \frac{P_{B,2}}{P_{B1}}= \frac{\sin(k_y(L_y+1)) - z \sin(k L_y)}{\sin(k_yL_y) - z \sin(k (L_y-1)} \xrightarrow[Ly\to\infty]{k_y = \pi + i \mu} -e^{\mu} .
\end{align}
This transcendental equation is similar to Eq.~\eqref{eq:consistency1} and depending on the value of $z$, now also a function of $\lambda$ and $k_x$, we loose a real-valued solution for $k_y$ at $k_y \approx \pi$. This solution is replaced by an evanescent solution with $k_y = \pi + i\mu$ and in the limit of $L_y \rightarrow \infty$, we obtain
\begin{align}
    \label{eq:consistencyL}
    e^{\mu} = -1/z = -\frac{\lambda - (-1 + t_1 + t_{+} e^{-ik_x} + t_- e^{ik_x})}{t_2}.
\end{align}
Then replacing $\cos(k) \rightarrow_{k\rightarrow \pi + i\mu} - \cosh(\mu) = \frac12 (z + 1/z)$ in the equation for $\lambda$~\eqref{eq:bulkspectrum}, we find the two solutions
\begin{align}
    \label{eq:SS2}
    &\lambda_{\chi = \pm} = h_0 + t_1 + \chi \sqrt{h_z^2 + t_2^2},
\end{align}
where $h_0(k_x)=-1 + (t_+ + t_-)\cos(k_x) $ and $h_z(k_x) = i(t_--t_+)\sin(k_x)$. These solutions coincide with the numerical spectrum in Supplementary Fig.~\ref{fig:boundaries}e. The edge mode dispersion, $\lambda_{\pm}$ is close to that in Eq.~\eqref{eq:lambdaSChi} but with a spectrum shifted by $t_1$. For some values of $k_x$ the spectrum of edge modes~\eqref{eq:SS2} hybridize with bulk modes. Compared to the boundary condition (1), this can be traced back to imposing detailed balance, which adds the $t_1$ term in \eqref{eq:boundA}, the sink potential.

\subsection{Topological properties}
\label{app:topological}

The probability distribution of active particles is determined by the non-Hermitian matrix $\hat{\bm W}_{\bf k}$ in Eq.~\eqref{eq:kolmbloch}. Non-Hermitian operators are classified according to their symmetries, to the gap structure of their complex energy spectra and to their dimension~\cite{Kawabata19}. Depending on the discrete symmetries of the system, and excluding crystal symmetries, the model falls in one of the 38 distinct topological classes that define strong topological insulators. The gaps are or three types: real or imaginary line gaps, or point gaps. If the class is topologically non-trivial for the given gap type and dimension, then a topological invariant exists to classify the possible topological phases of the system in that class. In our case, the matrix $\hat{\bm W}$ is non-Hermitian with time-reversal symmetry, $\hat{\bm W}^{*}_\mathbf{k}=\hat{\bm W}_{-\mathbf{k}}$, and inversion symmetry, $\hat{\sigma}_x\hat{\bm W}_{\bf k}\hat{\sigma}_x= \hat{\bm W}_{-{\bf k}}$.
It falls in the real Altland-Zirnbauer (AZ) symmetry class AI. Since $\hat{\bm W}$ has a point gap (see Fig ~\ref{fig:boundaries}) and describes a two-dimensional system, this class is trivial and has no strong topological invariant~\cite{Kawabata19}.

Our model can instead be better understood by analogy to weak topological insulators. Weak topological insulators in $d$ dimensions can be constructed by coupling $d-1$ strong topological insulators. Following this principle $\hat{\bm W}$ is constructed by coupling strong one-dimensional topological insulators defined in the $y$ direction  with non-reciprocal hoppings along the $x$ direction. In this way $\hat{\bm W}$ in a direct sum of two terms $\hat{\bm W}_{\bf k} = \hat{\bm W}_x + \hat{\bm W}_y$ with
\begin{align}
    \hat{\bm W}_x &= \left( -1 + (t_+ + t_-) \cos(k_x) \right) \hat{\sigma}_0 + i(t_- - t_+) \sin(k_x) \hat{\sigma}_z,\\
    \hat{\bm W}_y &= \left( t_2 + t_1 \cos(k_y) \right)\hat{\sigma}_x - t_1 \sin(k_y) \hat{\sigma}_y.
    \label{eq:directsum}
\end{align}
In these equations, $\hat{\bm W}_y$ describes a strong hermitian topological insulator in the $y$ direction. It is the Su, Schrieffer and Heeger model~\cite{Su79}. $\hat{\bm W}_y$ has chiral symmetry, represented by $\hat{\sigma}_z$ satisfying $\{\hat{\bm W}_y,\hat{\sigma}_z\} = 0$, and it is time-reversal symmetric, represented by complex conjugation. Therefore, $\hat{\bm W}_y$ belongs to the real AZ symmetry class BDI. Since $\hat{\bm W}_y$ is one-dimensional and has a real line gap (it is Hermitian), it can be classified by a strong topological invariant~\cite{Chiu161}, the total Berry phase on the Brillouin zone. It can be calculated for each sub-band $\eta = \pm$ in~\eqref{eq:eigenP} as~\cite{Ghatak2019_review}
\begin{align}
    \gamma_{y\eta} = \frac12\left(\gamma_{y\eta}^{LR} + \gamma_{y\eta}^{RL}\right) = \frac{1}{2\pi}\int_{-\pi}^{\pi} dk_y~ \partial_{k_y} \phi,
    \label{eq:totalBerry}
\end{align}
where we have introduced the Berry phase on the left ($P_{\eta}^{L}$) and right ($P_{\eta}^{R}$) normal modes of $\hat{\bm W}$
\begin{align}
    \gamma_{y\eta}^{\alpha\beta} = \frac{i}{2\pi} \oint_{\mathcal{C}} dk_y~ P_{\eta}^{\alpha*}({\bf k}) \partial_{k_y}P_{\eta}^{\beta}({\bf k}).
    \label{eq:windPdef}
\end{align}
In this last expression, the normal modes of $\hat{\bm W}$ are
\begin{align}
    P_{\eta}^{\alpha} = \frac{1}{\sqrt{1+e^{2i\alpha\theta}}} \left(
        \begin{array}{c}
            1\\
            \eta e^{i(\phi + \alpha \theta)}
        \end{array}
    \right),
\end{align}
with $\alpha = +1$ for right- and $\alpha = -1$ for left-eigenstates,  $\phi = {\rm arg}\left( t_2 + t_1\cos(k_y) + i t_1 \sin(k_y) \right)$, and $\theta = {\rm arg}\left( -\sum_{i} t_i + (t_++t_-)\cos(k_x) + i (t_- - t_+) \sin(k_x) \right)$.
The topological phase occurs when $t_1 > t_2$ because $\gamma_{y\pm} =1$, and the system is trivial otherwise, with $\gamma_{y\pm}=0$. This winding number is independent on $\hat{\bm W}_x$ in~\eqref{eq:directsum}, it is the same for both $\hat{\bm W}$ and $\hat{\bm W}_y$. Since the winding number $\gamma_{y\eta}$ defined in Eq.~\eqref{eq:totalBerry} is the winding number of the Hermitian operator $\hat{\bm W}_y$, we call it the Hermitian winding number, $w_{\rm H} = \gamma_{y}$.

The non-Hermitian matrix $\hat{\bm W}_x$ in Eq.~\eqref{eq:directsum} couples each copy of $\hat{\bm W}_y$ in the $x$ direction. Because $\hat{\bm W}_x$ is time-reversal symmetric and has no chiral symmetry, it can be classified in the real AZ symmetry class AI. Also, since it describes a one-dimensional system with an imaginary line gap, it can be classified by a topological invariant. As we now show, both the winding of the spectrum in the complex plane and the total Berry phase vanish for $\hat{\bm W}_x$, which is thus always trivial~\cite{Ghatak2019_review}. 
The reason the winding numbers of $\hat{\bm W}_x$ vanish can be derived from the fact that $\hat{\bm W}_x$ is composed of two independent copies, one for each eigenvalue $\chi = \pm$ of $\hat{\sigma}_z$,
\begin{align}
    H_{s, \chi}(k_x) = -1 + (t_+ + t_-) \cos(k_x) + \chi i(t_- - t_+) \sin(k_x).
    \label{eq:Hs}
\end{align}
These two copies are non-Hermitian matrices with the same symmetries than $\hat{\bm W}_x$ and each with opposite winding numbers of their spectrum in the complex plane~\cite{Okuma20,Ghatak2019_review,DasbiswasE9031}
\begin{align}
    w_{\mathrm{nH},\pm} = \frac{1}{2\pi i} \int_{-\pi}^{\pi} dk_x~ \frac{d \log\left(  H_{s,\pm}(k_x) - E \right)}{dk_x} = -w_{\mathrm{nH},\mp}.
    \label{eq:windE}
\end{align}
These winding number are $w_{\rm nH,\chi} = \chi (= \pm 1)$ for modes within an ellipse in the complex plane, centered at $\lambda_C = -1$ and with radii $t_++t_-$ on the real axis and $|t_+ - t_-|$ on the imaginary axis (the outer dashed black line in Fig.~4{f}). In general, these two winding numbers cancel out in $\hat{\bm W}$ since the two independent modes of $\hat{\bm W}_x$ are coupled by $\hat{\bm W}_y$ in the $y$ direction. As a consequence, the spectrum for periodic and open boundary conditions in the $x$ direction are similar to each other, so the non-Hermitian skin effect is absent (see Figs.~\ref{fig:boundaries} {b},{d}). 

{As derived in Sec.~\ref{app:edgemodes}, in the topological device a $y$ boundary that preserves chirality has two chiral edge modes, described by the two copies~\eqref{eq:Hs} on the top ($\chi = +$) and bottom ($\chi = -$) edges, since the chirality of $\hat{\bm W}_y$ is represemnted by $\hat{\sigma}_z$. In this case, each edge displays a one-dimensional first-order non-Hermitian skin effect because of their non-zero $w_{\mathrm{nH}}$, resulting in a second-order non-Hermitian skin effect of the two-dimensional system.}

Although the boundary condition that preserves chiral symmetry is a useful approximation, it is necessary to impose detailed balance at the top and bottom edge to obtain exactly the edge theory that we obtain numerically, as discussed in Sec.~\ref{app:edgemodes}. The main change between these two boundary conditions is that the spectrum of the edge modes shifts along the real axis, and as a result some of the edge modes partially hybridize with bulk modes~\eqref{eq:SS2}, see Supplementary Fig.~\ref{fig:boundaries}e. This effect can be interpreted as an edge potential on the topological edge states, which, despite breaking chiral symmetry, is not expected to change their skin-effect, for moderate parameter values. We confirm this expectation numerically, using the conditional probability parameters relevant for our experiments in Supplementary Fig.~\ref{fig:boundaries}f. 

With the above, our analysis implies that, if chiral edge modes exists with open boundary conditions in the $y$ direction (i.e. when $t_1/t_2>1$), these display a non-Hermitian skin effect with open boundary conditions in the $x$ direction, resulting in an accumulation of active particles at the corners. This effect is a second-order non-Hermitian skin-effect~\cite{Lee2019c,Ma2020,Okugawa20,Kawabata20b,Fu2020}, which exists when the topological number
\begin{align}
\label{eq:invariantapp}
    \nu = w_{\rm H} w_{\rm nH},
\end{align}
is non-zero. This is the case for our topological devices, but $\nu=0$ for our trivial devices. The Hermitian topology of $\hat{\bm W}_y$ when $w_{\mathrm{H}}\neq 0$ spatially separates the topological non-Hermitian modes of $\hat{\bm W}_x$ with opposite and non-vanishing $w_{\mathrm{nH}}$, realizing a second-order skin effect.

\subsection{Topological invariant of the second-order non-Hermitian skin-effect: connection to Hermitian topology}
\label{app:topological2}

By following Ref.~\cite{Okugawa20} it is possible to understand the second-order non-Hermitian skin effect by connecting with topological Hermitian systems as follows. First we construct an Hermitian matrix from $\hat{\bm W}$ as
\begin{equation}
   H_{\mathrm{eff}} = \left( 
        \begin{array}{cc}
            0 & \hat{\bm W}-\lambda_C\\
            \hat{\bm W}^{\dagger}-\lambda_C & 0
        \end{array}
    \right),
\end{equation}
where $\lambda_C$ is a base energy defined by the center of the point gap. This is in general a complex number, but, as can be seen from Supplementary Fig.~\ref{fig:boundaries}, in our case $\lambda_C=-1$.
When $\lambda_C$ is real, as in our case, it is instructive to perform a unitary transformation of $H_{\mathrm{eff}}$ as $\tilde{H}_{\mathrm{eff}}=U H_{\mathrm{eff}} U^{\dagger}$ by using the unitary matrix
\begin{equation}
   U = \left( 
        \begin{array}{cccc}
            0 & 0 & 0 & -1 \\
            1 & 0 & 0 & 0 \\
            0 & -1 & 0 & 0 \\
            0 & 0 & 1 & 0 
        \end{array}
    \right),
\end{equation}
such that
\begin{equation}
    \tilde{H}_{\mathrm{eff}} = \tilde{H}_x \otimes \tau_0 + \sigma_0 \otimes \tilde{H}_y
\end{equation}
\begin{align}
    \tilde{H}_x &= \left( -1-\lambda_C + (t_+ + t_-) \cos(k_x) \right)\hat{\sigma}_x - (t_- - t_+) \sin(k_x) \hat{\sigma}_y,\\
    \tilde{H}_y &= \left( -t_2 - t_1 \cos(k_y) \right)\hat{\tau}_x - t_1 \sin(k_y) \hat{\tau}_y.
    \label{eq:directsumH}
\end{align}
Both $\tilde{H}_x$ and $\tilde{H}_y$ have chiral symmetry represented by $\hat{\sigma}_z$ and $\hat{\tau}_z$, respectively. Each symmetry is associated to the winding numbers $\tilde{\omega}_{x,y}$. 
Consequently, $\tilde{H}_{\mathrm{eff}}$ has the symmetry $\tilde{\Gamma} = \hat{\sigma}_z\otimes \hat{\tau}_z$.
Additionally it displays inversion symmetry given by $\tilde{I}= \hat{\sigma}_x\otimes \hat{\tau}_x$. 
Note that $\tilde{\Gamma}$ commutes with  $\tilde{I}$.

As discussed in Sec~\ref{app:topological}, the present model has trivial non-Hermitian topology in the bulk but non-trivial on the edges. In the nomenclature of Ref.~\cite{Okugawa20} its second-order non-Hermitian skin effect is thus extrinsic, since chirality and inversion symmetries commute and the non-Hermitian topology is only characterized by chiral symmetry  $\tilde{\Gamma}$. 
The corresponding Hermitian winding numbers of $\tilde{H}_x$ and $\tilde{H}_y$ are non-trivial and can be computed from~\eqref{eq:totalBerry}. We have that 
\begin{enumerate}
    \item the normal modes of $\tilde{H}_x$ are
    \begin{align}
        P_{\eta = \pm} = \frac{1}{\sqrt{2}}
        \left(
        \begin{array}{c}
            1\\
            \eta e^{i\phi_x}
        \end{array}
        \right)
    \end{align}
where $\phi_x = {\rm arg}\left[ -(1+\lambda_C) + (t_++t_-)\cos(k_x) + i (t_+ - t_-)\sin(k_x) \right]$. So the winding number $w_x$ is 
    \begin{align}
        w_{x} = \frac{1}{2\pi}\int_{-\pi}^{\pi} dk_x \partial_{k_x} \phi_x.
    \end{align}
    $w_{x}= 1$ for $\lambda_C \in -1 \pm (t_+ + t_-)$ and it vanishes otherwise.
    \item the normal modes of $\tilde{H}_y$ are
    \begin{align}
        P_{\eta = \pm} = \frac{1}{\sqrt{2}}
        \left(
        \begin{array}{c}
            1\\
            \eta e^{i\phi_y}
        \end{array}
        \right)
    \end{align}
    where $\phi_y = {\rm arg}\left[ -(t_2+t_1\cos(k_y)) - i t_1 \sin(k_y) \right]$. So the winding number $w_y$ is
    \begin{align}
        w_{y} = \frac{1}{2\pi}\int_{-\pi}^{\pi} dk_y \partial_{k_x} \phi_y.
    \end{align}
    $w_{y}= 1$ for $t_1/t_2 > 1$, which defines the topological device, and it vanishes in the trivial device.
\end{enumerate}
Comparing to our discussion in the previous section, we see here that the Hermitian winding number, $w_{\rm H}$, of $\hat{\bm W}$ coincides with the total Berry phase $w_y$ of $\tilde{H}_y$. Moreover, the non-Hermitian winding number, $w_{\rm nH}$, coincides with the total Berry phase $w_x$ of $\tilde{H}_x$ for eigenvalues $\lambda_C$ on the real axis, as in our case. The topological invariant we introduced in the main text Eq.~(3) (Eq. \eqref{eq:invariantapp} in this Supplementary Material) thus coincides with the one proposed in Ref.~\cite{Okugawa20}
\begin{align}
    \nu = w_{2D} \equiv w_x w_y, 
\end{align}
for $\lambda_C \in \mathbb{R}$. Note that if we had considered $\lambda_C \in \mathbb{C}$, then there would be an additional chirality-breaking contribution $-{\rm Im}[\lambda_C]\hat{\sigma}_2\otimes\hat{\tau}_3$ that enters~\eqref{eq:directsum}, that would have prevented this topological classification~\cite{Okugawa20}.

\subsection{Discussion on the topological invariant $\nu$}
\label{app:discussionnu}

In Sections~\ref{app:topological} and \ref{app:topological2} we computed the same topological invariant, $\nu$, from two different arguments. One may wonder if this invariant is truly a 2D invariant. First, we note that $\nu$ is closer to invariants that characterize weak topological insulators, as it is defined by a combination of lower dimensional (in our case 1D) topological invariants. In this sense, the topological invariant describing our system is not a bulk 2D topological invariant. Rather, the present higher-order topological behaviour emerges from the superposition of Hermitian and non-Hermitian 1D topologies in transverse directions, that combined lead to a second-order non-Hermitian skin-effect. 

We note as well that $\nu$ is not strictly a 1D invariant either since a first order non-Hermitian skin-effect cannot occur in the presence of inversion. As discussed in the main text, in our experiments we enforce detailed balance over the unit cell, so the lattice does not break inversion symmetry, and thus does not allow a first-order skin-effect. Each edge independently may break 1D inversion symmetry, and cause the particle accumulation. The parameters for which our 2D model allows this to happen are captured by $\nu$.

\section*{Supplementary Discussion 2 : Comparisons between devices with different dimensions and densities}
\label{app:sum}

In this section we review all our results for the trivial and topological devices, in small ( $(L_x,L_y) = (12,6)$ ) and large ( $(L_x,L_y) = (13,14)$, main text) devices (see Supplementary Fig.~\ref{fig:review} a-d) and for low and high density of active particles (see Methods, Experimental Setup). 

\subsection{Topological chiral edge motion}

We initiate each device with particles on the top left corner and compare our experimental data with our model in Supplementary Fig.~\ref{fig:review} e. In order to decrease the statistical uncertainty, we use the symmetry between opposite corners to include trajectories from the bottom right corner in the figures. 

At long times, fewer particles contribute to the average because some have shorter trajectories than others. The agreement between model and theory is generally better for shorter times because the experimental average then includes more particles, from 150 to 300 particles. 

\subsection{Non-Hermitian topology : corner localization}

We initiate each device with a uniform density of particles on every cell. The time evolution of the density density resembles that in Figs.~4 a,b of the main text, drawn for the large device with a large density of active particles, resulting in an accumulation of active particles at the corners. This increase in the density of particles at the corners is compared quantitatively using the Shannon entropy, in Supplementary Fig.~\ref{fig:review}g. 

Because the entropy is sensitive to the noise in the density of particles, we reduce it by averaging the density over adjacent cells. The range for this averaging is estimated by a Fourier transform, to determine how much the distribution of particles spreads away from uniformity. We obtain the spread $\Delta k$ from which we deduce the range of fluctuations $\Delta r = 1/(2\Delta k)$. We then perform the averaging procedure over $3\Delta r \approx 3$ cells for the small devices, and $\approx 2$ cells for the large devices. We perform this averaging procedure on both experimental and theoretical data.

The entropy we compute from our model is usually larger than what is observed experimentally. This can be caused by the residual noise from the fluctuations in the number of particles or by correlations between particles, that tend obstruct the motion of each other and lead to jamming. Also we see a difference between the entropies of the topological and trivial devices when the density of active particles is large. In this case the larger number of particles per cell enables the use of the entropy as a probe of the non-Hermitian skin effect.

\section*{Supplementary References}

\end{document}